\documentclass[prd,aps,amssymb,amsmath,tightenlines,nofootinbib]{revtex4}
\usepackage{graphicx,epsfig}
\usepackage{slashed}
\usepackage[colorinlistoftodos]{todonotes}
\usepackage{appendix}

\begin{document}

\leftline{KCL-PH-TH/2021-{\bf 80}}
\vspace{1cm}

\title{Torsion in string-inspired cosmologies and the universe dark sector}

\author{Nick E. Mavromatos$^{a, b}$}

\affiliation{$^a$ Department of Physics, School of Applied Mathematical and Physical Sciences, National Technical University of Athens, 9 Iroon Polytechniou Str., Zografou Campus, 15780 Athens, Greece \\
$^b$Theoretical Particle Physics and Cosmology Group, Department of Physics, King's~College~London, Strand, London WC2R 2LS, UK}

\begin{abstract}

\vspace{0.2cm}

I review several aspects of torsion in string-inspired cosmologies. In particular, I discuss its connection with fundamental, string-model independent, axion fields
associated with the massless gravitational multiplet of the string. I argue in favour of
the role of primordial gravitational anomalies coupled to such axions in inducing inflation of a type encountered in the ``running-vacuum-model (RVM) '' cosmological framework, without fundamental inflaton fields. The gravitational-anomaly terms owe their existence to  the Green-Schwarz mechanism for the (extra-dimensional) anomaly cancellation, and may be non-trivial in such theories in the presence of (primordial) gravitational waves at early stages of the four-dimensional string Universe (after compactification). I also discuss how the torsion-induced stringy axions can acquire a mass in the post inflationary era, due to non perturbative effects, thus having the potential to play the r\^ole of (a component of) dark matter in such models. Finally, I describe briefly the current-era phenomenology of this model, with the emphasis placed on the possibility of alleviating tensions observed in the current-era cosmological data. A brief phenomenological comparison with other cosmological models in contorted geometries is also made.
 
\end{abstract}

\maketitle

\section{Introduction: A case for going beyond GR and Riemannian Geometry \label{sec:intro}}

General Relativity (GR) has provided a very successful framework for the classical description of the gravitational interaction~\cite{iorio}, which works and tested very well at both local (terrestrial, astrophysical) and global (cosmological)  scales.  Local GR is tested very successfully with solar data~\cite{will}.  
At large, cosmological scales, the global version of GR, namely its Friedmann-Lemaitre-Robertson-Walker (FLRW) cosmological solution, in the presence of a positive cosmological constant, $\Lambda > 0$, including a Cold-Dark-Matter component ($\Lambda$CDM paradigm), seems to describe well the precision measurements in modern Cosmology, coming from a plethora of diverse observations~\cite{snIa,snIa2,Planck}. 

The $\Lambda$CDM paradigm
 changed our perception for the evolution and properties of the Universe that we have held for most of the 20th century. According to the overwhelming majority of the data interpretations, the Universe appears to be currently accelerating, entering again a (approximately) de Sitter-like phase, with an equation of state almost that of a positive-cosmological-constant-dominated universe $w\simeq - 1$, like in the inflationary era, but with a much-smaller-in-magnitude Hubble parameter. Ordinary (mostly baryonic) matter constitutes
{\it only} $\sim 5\%$ of the current-epoch energy budget of the Universe. The remaining 95\% of the observable Universe current energy budget consists of unknown substances: $\sim 25\%$ of the Universe budget appears to be dark matter (DM), which currently is of unknown origin, and we only have astrophysical (i.e. gravitational) evidence for its existence, and $\sim 70\%$ corresponds to an unknown form of energy (dark energy), responsible for the Universe acceleration, which in the current-era behaves as a positive cosmological constant.\footnote{There is, however, the trend of modified Newtonian dynamics (MOND)~\cite{mond}, and its relativistic field theory variants~\cite{teves}, which claim that there is no DM in the Universe, but only modified gravity laws at galactic scales. At present, though, such alternative to DM models appear to be in significant tension with data from colliding galaxies, which seem to favour the DM hypothesis, including DM self-interactions~\cite{coll}. However, such issues are still open for debate, and will not be discussed further here.}

In 2015, the discovery of gravitational waves (GW) by the LIGO/VIRGO interferometers~\cite{ligo} 
has given another great boost to GR, as it 
confirmed one of its greatest predictions. The waves were the result of the merging of two celestial objects, which have been claimed to be rotating black holes. This discovery opened up a new era for astronomy, which can be used to test further GR, in the sense of potentially constraining the parameters of, or falsifying, modified gravity theories beyond GR, where, e.g. black holes can exist characterised by (scalar-field) secondary hair. 
Moreover, imaging of the shadow of a supermassive black hole in the centre of the giant elliptical galaxy M87 in the Virgo constellation by the Event-Horizon Telescope (EHT)~\cite{shadow} is expected to contribute novel, and in some respects complementary to GW, ways of constraining black hole physics~\cite{psalt}, and in general testing beyond-GR classical gravity theories.  Indeed, black-hole solutions in modified gravity theories would have observable (in principle) modifications of their shadows compared to those of the GR black holes.  One of the types of such black holes arises in string-inspired modifications of GR, which involve  exponential couplings of a scalar (dilaton) field with quadratic-in-space-time-curvature-tensors Gauss-Bonnet (GB) combinations~\cite{bhscal}. In recent years, such modifications of GR have also been considered in the framework of the so called scalar-tensor (Horndeski-type) gravity theories, where linear couplings of the scalar field with GB terms occur, respecting a shift symmetry amounting to constant shifts of the scalar~\cite{horn,horn2,horn3,horn4,horn5,horn6,horn7}. These models also have black holes with secondary scalar hair. Other theories beyond GR, whose black hole solutions can be probed experimentally, are the so called Chern-Simons theories of gravity~\cite{jackiw,chern}, which involve CP violating anomalous derivative couplings of pseudoscalar (axion-like) fields with gravitational Chern-Simons terms. 

However, despite its phenomenological successes,  the need to go beyond GR is rather pressing from a theoretical point of view.  {\it Quantization of Gravity}, that is the development of {\it both} a mathematically {\it and} physically consistent theory of {\it Quantum Gravity} (QG), following the example of the rest of the fundamental interactions in Nature, still remains an important outstanding question in fundamental physics, which, if resolved, will undoubtedly lead to concrete modifications of GR~\cite{QG}. GR as it stands is a non-renormalizable theory, and, thus, it cannot, by itself, be the appropriate theory for QG. \footnote{Whether gravitational interactions, though, have to be quantised like the other interactions in Nature is still an open issue. For instance, in \cite{verlinde} gravity is conjectured to be an entropic
force, which has its origin in changes in the information associated with the positions of material bodies. In this review we shall not adopt this viewpoint, but assume that gravity needs to be quantised.} Although effective low-energy field theories, based on GR, of a quantised weak gravitational field about a given space-time background are currently in operation, 
and may give important information on the (perturbative) fixed-point structure of gravity (see, {\it e.g.} the ``asymptotic safety'' approach~\cite{fixed}), they do have 
limited success, since they cannot provide answers to the fundamental question on the dynamical emergence of space time itself. 

At present there are several frameworks for dealing with such fundamental QG issues. The most developed is 
String theory~\cite{string,string2}, which offers a consistent quantization of gravity, along with the other fundamental interactions. This framework leads to concrete modifications of GR, given that the low-energy gravitational effective actions contain (an infinite number of) higher-order terms in both derivatives and curvature tensors. Nonetheless, at present, string theories do not constitute a truly background independent formalism, although the potential for that may exist in future non-perturbative formulations. Indeed, all known formulations of string theory to date, including D-brane theories, appear to be space-time background dependent in the sense that the emergence of the space-time itself is still not completely understood at the truly microscopic level. Moreover, string theory is plagued by the plethora of allowed vacua, the so-called Landscape, and the mechanism for the choice of the physical vacuum is still not understood (some physicists even invoke anthropic arguments to explain such a choice~\cite{landscape}). 

Mathematically and physically consistent {\it background-independent} QG frameworks, alternative to string theory,  which have the potential of describing dynamically the appearance of space time, do currently exist, and include: Loop Quantum Gravity (LQG)~\cite{loop,loop2} and its application to homogeneous systems, the so-called Loop-Quantum Cosmology~\cite{LQC,LQC2}, the spin-foam models~\cite{foam,foam2} and 
the Group-Field-Theory approach to QG~\cite{gftoriti}, which defines quantum field theories {\it of spacetime} in which the base manifold is an appropriate Lie group, describing the dynamics of both the topology and the geometry of the universe. The perturbative limit of LQG gives rise to spin-foam models. Spin-foam models corresponding to finite groups also exist in the contemporary literature~\cite{BDR}.

Apart from the theoretical needs to go beyond GR~\footnote{At this point, I should stress, of course, that 
although from a point of view based on String Theory  (or traditional Quantum-Field-Theory), significant modifications of GR might be expected in a successful quantum-gravity theory, this may not be the case in the other approaches, such as LQG, whose 
philosophy is more about changing the way the quantum theory is built, so as to be defined non-perturbatively. Nonetheless, in this review I will concentrate on versions of LQG which go beyond Riemannian geometry, by being characterised by non-trivial torsion, and in this respect there are notable deviations from GR in the respective continuous low energy limits, as we shall discuss.}, there seem to be observational issues as well, which prompted several theorists to seek models that go beyond the $\Lambda$CDM paradigm of cosmology, and, more generally, beyond GR. 
 Indeed, the cosmological data appear to be characterised by interesting tensions at small scales. We have, for instance, the so-called ``$H_0$-tension'',  pointing to discrepancies between the current-epoch value of the Hubble parameter $H_0$, as  inferred from local (Cepheid galaxies) measurements~\cite{H0,H02,H03}, and  that obtained by means of cosmic-microwave-background (CMB) measurements of the Planck collaboration, based on a fits within the $\Lambda$CDM framework~\cite{Planck}. 
Current structure-formation data also seem to be characterised by a disparity in the root-mean-square (r.m.s) value of the current-epoch matter density fluctuations, within spheres of radius 8$h^{-1}$ (with $h \simeq 0.7$ the ``reduced Hubble constant''),  inferred by $\Lambda$CDM fits, as compared to the r.m.s. value obtained by local direct measurements at low redshifts~\cite{sigma8,sigma82,sigma83}. This latter type of tension is called the ``$\sigma^8$ tension''. Although one cannot yet exclude the possibility that  such tensions are due to insufficient (at present) data accuracy and/or statistics, given that the pertinent discrepancies  are all currently  within $2\sigma-3\sigma$, nonetheless they caused excitement among the pertinent physics communities and their resolution prompted research into theoretical models beyond $\Lambda$CDM, including going beyond GR, via modified gravity theories~\cite{mena,perivol}. We mention at this point that one such theoretical framework  is the so-called ``running vacuum model (RVM)'' framework~\cite{sola,sola2,sola3} (where a time dependent vacuum energy is assumed, but with equation of state $w_{\rm vacuum} = -1$ as in the de Sitter case). This provides an effective smooth evolution of the Universe, from inflation till the current era~\cite{lima,perico,baslima,baslima2}, with, in principle observable deviations from $\Lambda$CDM~\cite{sola5,sola6,sola7,sola8,sola9,tsiapi1,tsiapi2}, which notably can also alleviate simultaneously both types of tensions, $H_0$ and $\sigma_8$, in the current data~\cite{sola10,sola11}.\footnote{For another model with dynamical, ``running'', dark energy, where the space-time is associated with a dynamical field (dynamical space time), and its phenomenological consequences, see \cite{benisty,benisty2}.}    
In this review we shall derive such an RVM cosmology as an appropriate  low-energy limit of string theories, under some special circumstances that we shall discuss in detail~\cite{sola4,ms1,ms1b,ms2}.
 
The emergent space-time geometries in all the above approaches, including string theory, need not be Riemannian. Indeed,
torsion~\cite{torsion} could well be present in the effective gravitational field theories arising from the above microscopic models of QG. For instance, generalised geometries with torsion may arise in the low-energy limit of string theory~\cite{string,string2}, where the torsion is provided by the field strength of the spin-one antisymmetric tensor  (Kalb-Ramond (KR)) field of the massless bosonic gravitational multiplet of strings.
In this case, the torsion has only a totally antisymmetric component in its world indices, due to the respective total antisymmetry of the field strength. In fact, in the effective four-dimensional gravitational field theory, which is obtained after compactification of the extra-dimensional string theory,  this torsion is equivalent in the full quantum theory to a dynamical massless pseudoscalar  field (KR axion), which is identified with the so-called string-model independent axion~\cite{kaloper,svrcek}. The presence of the pseudoscalar leads also to axion secondary hair in the (spinning) black holes of the theory~\cite{kaloper}.\footnote{At this point we mention that there are several interesting works in the literature dealing with the r\^ole of string-inspired KR fields in cosmology, for instance, in the context of modified gravity models we refer the reader to the works in~\cite{tanmoy1,tanmoy2,tanmoy3,tanmoy4} for further details and comparison with our results reported in this review. For brane-world models, see, {\it eg.} Ref.~\cite{chak}, whilst for a potential r\^ole of KR-inspired torsion (as an axion field) in inflationary magnetogenesis in Chern-Simons electrodynamics see Ref.~\cite{chak2}.}
It worths remarking that the association of the totally antisymmetric component of the torsion with an axion field seems to be a generic property of also contorted field theories with fermions, independent of string theory, for instance contorted quantum electrodynamics (QED)~\cite{kaloper}, which we discuss in the Appendix of this review, as a prototype. Indeed, coupling Einstein-Cartan theory ({\it i.e.} a pure gravitational theory with a scalar curvature term in the Lagrangian, in the Palatini formalism, where vielbeins and spin connections are viewed as independent fields, having removed the zero-torsion constraint) to Dirac fermion fields, is the simplest theory where the equations of motion for the torsion become non trivial. 
In the context of contorted LQG,  the torsion couples to the so-called Barbero-Immirzi (BI) parameter~\cite{bi,bi2},\footnote{The BI parameter arises in the framework of LQG when one attempts to express the Lorentz connection of the non compact group SO(3,1) in terms of a complex connection which takes on values in a compact group of rotations (SO(3) or its double cover SU(2)). This parameter measures the quantum of area in LQG, and therefore plays an important r\^ole in black hole thermodynamics. Its value can be fixed by matching the semi-classical entropy of a black hole with the counting of 
microstates within the LQG framework.} $\beta$, which appears as a coefficient of the Holst term in the effective (continuous) gravitational action~\cite{torsion2}: $\frac{1}{2\kappa^2 \, \beta}\, \int d^4 x \sqrt{-g}\, \varepsilon^{\mu\nu\rho\sigma} \, \widehat R_{\mu\nu\rho\sigma}$, where $\kappa = \sqrt{8\pi\, {\rm G}} = M_{\rm Pl}^{-1}$, with $M_{\rm Pl} =  2.43 \times 10^{18}$~GeV  (we work in units of $\hbar=c=1$ throughout), is the four-dimensional gravitational constant, 
$ \varepsilon^{\mu\nu\rho\sigma}, \, \mu, \nu, \rho, \sigma =0, \dots 3,$ is the gravitationally covariant Levi-Civita antisymmetric tensor density and $\widehat R_{\mu\nu\rho\sigma}$ is a curvature tensor with torsion~\cite{torsion} (for notation and conventions see Appendix). In the absence of torsion, the BI parameter $\beta$ does not play any r\^ole, due to the cyclic property of the conventional Riemann tensor, which implies  
$\varepsilon^{\mu\nu\rho\sigma} \, R_{\mu\nu\rho\sigma} =0$. But obviously, this is not the case in contorted geometries. 
 The BI parameter also appears in Lorentzian spin-foam models of QG, which are defined on an appropriate  path integral over {\it discrete} Lorentzian quantum geometric configurations (``{\it geometry quanta}"), which include metric and torsion degrees of freedom. The torsion degrees of freedom arise due to an anomaly of the algebra of the constraints, imposed on the fundamental geometry quanta, which is parametrized by the BI parameter.
 
In such theories, the BI parameter and torsion affect the gravitational dynamics, leading to in-principle observable effects~\cite{torsion2,shapiro,andrade1}. Cosmologies with torsion~\cite{tors3,tors2}, including modified contorted gravity theories, e.g. $f(\widehat R)$ theories~\cite{tors1}, have been extensively considered in the literature, and their phenomenology has been studied in detail, including a possible relation of the matter-antimatter asymmetry and the dark sector of the universe with fermionic-torsion condensates~\cite{popl1,popl2}, arising from the four-fermion interactions that are characteristic of all contorted theories in the presence of fermions. At this point we should also mention the so-called {\it teleparallel} theories of gravity~\cite{tele,heis1,heis2,heis3,tele2}, which, in contrast to the aforementioned contorted theories of gravity, in which both the metric and the torsion fields co-exist, are only characterised by the presence of the torsion field, which mimics the dynamics of the gravitational field. For some interesting cosmic dynamo effects in teleparallel theories and connection with gravitational anomalies the reader should also consult~\cite{andrade,andrade2}. 
For further investigations on such generalisations of GR, based on teleparallel geometries see Refs.~\cite{heis4,heis5}, where teleparallel models beyond $f(T)$ (with $T$ denoting the torsion scalar)~\cite{tele}, including the so-called non-metricity scalar $Q$ ($f(Q)$ gravity), and their applications to cosmology and black-hole physics are considered. 

Torsion constitutes therefore a rather vast research area, rich in physical applications (albeit not yet confirmed experimentally !), as, we hope, becomes clear from the above introduction, and hence we cannot review all of its aspects in this short review. Here we shall focus instead only on a particular kind of torsion, already mentioned previously, namely the one which stems from the antisymmetric tensor field of the bosonic massless gravitational multiplet of the string~\cite{string,string2,kaloper}, and shall discuss its connection with the dark sector of the Universe in the context of a low-energy string-inspired cosmology with non-trivial gravitational anomalies~\cite{anom} at early epochs of the universe~\cite{sola4,ms1,ms1b,ms2}. The anomalous terms are expressed by the presence in the Lagrangian of a  CP-violating gravitational Chern-Simons term~\cite{jackiw,chern} coupled to the string-model independent KR axion, which in (3+1)-dimensional space times is dual to the torsion in a way we shall explain in detail. The anomaly terms are remnants in four dimensions of the higher-dimensional counterterms that are inserted in the theory as a consequence of the anomaly-cancellation mechanism of Green-Schwarz~\cite{string,string2}. The gravitational anomalies are fully consistent with the general covariance of the effective action, but describe a non-trivial exchange of energy between the axion and gravitational fields. These gravitational Chern-Simons terms are non-trivial when there are (primordial) gravitational waves (GW) present~\cite{stephon,sola4}, whose potential origin has been discussed in \cite{ms1,ms1b}, and will also be reviewed briefly in the present work. 
This interaction between the KR axion and gravitational Chern-Simons terms proves essential in leading~\cite{sola4,ms1,ms1b} to a cosmological vacuum of the RVM type~\cite{sola,sola2,sola3}, which in turn implies an early inflationary era, without inflaton fields~\cite{lima,perico,baslima,baslima2}. 
The formation of GW condensates is crucial to the effect. Moreover, such condensates lead to undiluted KR axion backgrounds at the exit from inflation, of a form which break  {\it spontaneoysly} Lorentz symmetry. These backgrounds 
survive well into the radiation era~\cite{sola4,ms1,ms1b} and lead to leptogenesis in theories with right-handed massive sterile neutrinos~\cite{decesare,bms1,bms2,bms3,bms4}. In the post inflationary epoch, non-perturbative effects, during the QCD era, can generate, under some circumstances, a potential and a mass for the KR axions~\cite{ms2}, which thus has the potential of playing the r\^ole of (a component of) DM in this string-inspired universe. Due to the link of the KR axion with torsion, then, one could have a {\it geometrical origin} of DM in such models. The RVM nature of this  effective string-inspired cosmological model is maintained until the current era~\cite{sola4}, and in this sense the model has observable in principle variations from the $\Lambda$CDM, contributing in parallel to the alleviation of the cosmological tensions in the data~\cite{sola5,sola6,sola7,sola8,sola9,sola10,tsiapi1,tsiapi2,sola11}. 

The structure of the review is the following: in the next section \ref{sec:rvm}, we review briefly the essential features and underlying formalism of the RVM vacuum, in order to introduce the reader into the basic concepts and techniques, that we shall make use of in  subsequent parts of the article. In section \ref{sec:string}, we present the string-inspired (3+1)-dimensional gravitational model, and explain the appearance of torsion and its equivalence to a dynamical massless KR axion field. A discussion on the origin and properties of the crucial anomalous gravitational Chern-Simons terms in the effective action, and a demonstration of their consistency with general covariance, is given. 
In the following section \ref{sec:gw}, we discuss briefly the origin of primordial gravitational waves (GW) and their r\^ole in inducing non-trivial gravitational anomaly terms. We evaluate  the appropriately induced anomaly condensates, and demonstrate the spontaneous breaking of Lorentz invariance  by the vacuum of the theory due to the presence of Lorentz-Invariance-Violating (LIV) non-trivial KR axion backgrounds. We explain carefully, under which conditions such condensates lead to dynamical RVM-type inflation without inflaton fields. In section \ref{sec:postinfl} we discuss the post inflationary era, where gravitational anomalies are cancelled by the generation of chiral fermion matter at the exit from inflation, in a way we shall explain in detail. We describe briefly the non-perturbative mechanism by means of which a mass can be generated for the KR axion fields, which can thus play the r\^ole of (a component of) DM in the effective cosmological model.
We also discuss briefly how leptogenesis occurs in models involving sterile massive right-handed neutrinos, which can be  accommodated in our string-inspired framework. Finally, in section \ref{sec:modern}, instead of conclusions, we discuss briefly the modern era, in which the RVM form of the vacuum of our cosmological model is maintained, leading to observable in principle modifications of $\Lambda$CDM and tension alleviations. We also make a brief phenomenological comparison of our string-inspired torsionful cosmology with some other cosmological models in contorted geometries that exist in the contemporary literature. Some mathematical properties of torsion, that we make use of in the article, are outlined in the Appendix, in the context of an instructive example, that of contorted quantum electrodynamics with Dirac fermions coupled to an Einstein-Cartan gravitational sector~\cite{kaloper}. We also make a comparison there with other topological modifications of contorted gravity existing in the contemporary literature.

\section{Review of the Running Vacuum Model (RVM) framework \label{sec:rvm}}

The Running Vacuum Model (RVM) of the Universe is a phenomenological framework~\cite{sola,sola2,sola3}, 
which postulates that the cosmological vacuum is characterised by a de-Sitter-type equation of state, but the vacuum energy density depends on the cosmic time, through its dependence on the Hubble parameter $H(t) = \frac{\dot a}{a}$, in the Robertson-Walker (RW) frame (where $a=a(t)$ is the scale factor of the FLRW universe, assumed {\it spatially flat}, as suggested by the data~\cite{Planck}, and the overdot over a quantity denotes its cosmic-time derivative $d/dt$):
\begin{align}\label{rvmeos}
p_{\rm RVM} (H(t)) = - \rho_{\rm RVM} (H(t)).
\end{align} 
For reasons of general covariance~\cite{sola,sola2,sola3}, the energy density of the vacuum is a function of even powers of $H(t)$. 
In fact the phenomenological RVM framework postulates a renormalisation-group-like equation for the vacuum energy density~\cite{sola,sola2,sola3} 
\begin{align}\label{rge}
\frac{d \,\rho_{\rm RVM} (H(t))}{d\,{\rm ln}(H^2(t))} =  \sum_{n=1}^\infty a_{2n} \, H^{2n}(t) 
\end{align} 
in the RW frame. 
  
In general, one may have additional dependence on $\dot  H(t)$. However, for our purposes here we make the phenomenologically viable assumption that within a particular era of the universe, the deceleration  parameter of that era $q = -(a \ddot a)/(\dot   a)^2$
is {\it approximately constant}, and thus 
$\dot H $ can be approximately expressed as $H \simeq - (1 + q) H^2(t)$, hence the dependence only on $H^{2n}$ 
in \eqref{rge} suffices to explain in a smooth way the cosmological evolution of our Universe from inflation till the present era~\cite{lima,perico,baslima,baslima2}.  It turns out that, for phenomenological reasons, only terms up to $H^4$ in \eqref{rge} play a r\^ole in the entire history of our Universe  from the inflationary era until today. Hence, solving in this case \eqref{rge}, yields
for the RVM vacuum:
\begin{align}\label{rgeh4}
\rho_{\rm RVM} = a_0 + a_2 \, H^2(t) + a_4 H^4(t) + \dots \equiv  
\frac{3}{\kappa^2} \, \left(c_0 + \nu H^{2} + \alpha
\frac{H^{4}}{H_{I}^{2}} + \dots \right) \, >\, 0\;,
\end{align}
where in the second equality we used the standard notation of the RVM. The quantity $H_I$ is the inflationary scale, whose value can be inferred from the recent Planck Collaboration data~\cite{Planck}, 
\begin{align}\label{PlHI}
 \frac{H_I}{M_{\rm Pl}} \simeq 10^{-5},
 \end{align}
and $c_0 \ge 0 $,  $\nu$ and $\alpha$ constants. In the conventional RVM, $\nu > 0$ and $\alpha >0$, and they are assumed constants through the entire evolution of the universe. The constant $c_0$ plays the r\^ole of the positive cosmological constant which is believed to be responsible for the current-era acceleration of the universe, according to the 
$\Lambda$CDM paradigm~\cite{Planck}. Its value is just an integration constant stemming from integrating Eq.~\eqref{rge}, and thus cannot be determined in this phenomenological effective framework.
The coefficients $\nu \ll 1$, as required by phenomenology (see discussion below), while $\alpha $ can be of order $\mathcal O(1)$.

It is not a trivial task to derive \eqref{rgeh4} from microscopic quantum field theories in curved space. In fact, this is highly model dependent. For instance, within simple quantum field theory models~\cite{qft,qft2}, the $H^4$ term cannot be derived, only the $H^2$ is. In this review we shall argue on the derivation of the $H^4$ term as well, which is responsible for inducing inflation, in a specific string-inspired model with gravitational anomalies, and the term $H^4$ is precisely due to anomaly condensates in the presence of GW perturbations~\cite{sola4,ms1,ms1b}, and is not related to ordinary matter effects. Moreover, as discussed in detail in \cite{sola4}, and will be reviewed below, the coefficients $\nu$ and $\alpha$ are not constant through the evolution of the universe, but their values depend on the era. During the inflationary era, for instance, the coefficient $\nu < 0$, as a result of contributions from the gravitational Chern-Simons anomalous terms, 
while it is positive in post inflationary eras, including the current one, due to different contributions it receives then from cosmic electromagnetic fields. 

Before moving onto the string-inspired cosmological model, let us first review briefly how the RVM frameork \eqref{rge} describes the entire evolution of the universe from its early inflationary de-DSitter era, to the modern one, where again the universe seems (from the current interpretation of cosmological data) to enter another de Sitter phase. This comes about by considering the Einstein's equations written in the form:
\begin{align}\label{einst}
R_{\mu\nu} - \frac{1}{2} g_{\mu\nu} R = \kappa^2  \Big(T_{\mu\nu}^{m} + T_{\mu\nu}^{\rm RVM} \Big)
\end{align}
where the subscript ``$m$" (``RVM") denotes matter/radiation (running vacuum) contributions to the stress tensor 
$T_{\mu\nu}$. From the Bianchi identity for the curvature tensor $\Big(R_{\mu\nu} - \frac{1}{2} g_{\mu\nu} R\Big)^{;\,\nu} =0$,
where the symbol $;$ denotes gravitational covariant derivative (in a torsion-free space-time, like the ordinary FLRW universe of interest here), and using \eqref{rgeh4} in the expression of the temporal components of \eqref{einst}, and the Freedman equation $\kappa^2 \, \rho_m + \Lambda (H) = 3\,H^2 $, 
we easily arrive at the evolution equation for the Hubble parameter in the RVM universe~\cite{lima,perico,baslima,baslima2}:
\begin{align}\label{evolH}
\dot H + \frac{3}{2} \, (1 + \omega_m) \, H^2 \, \Big( 1 - \nu - \frac{c_0}{H^2} - \alpha \, \frac{H^2}{H_I^2} \Big) =0,
\end{align}
with $w_m = \rho^m/p^m$ the equation of state of matter/radiation (assumed to be an ideal fluid with energy density $\rho^m$ and pressure $p^m$). Assuming constant $c_0, \nu, \alpha$ throughout the universe evolution, we obtain from
\eqref{evolH} the following general solution for $H(t)$:  
\begin{align}\label{evolHsol}
 H(a) = \Big(\frac{1 - \nu}{\alpha}\Big)^{1/2} \, \frac{H_I}{\sqrt{D \, a^{3(1-\nu)(1 + w_m)} + 1}},
 \end{align}  
 with $D > 0$ an appropriate integration constant, to be fixed phenomenologically.
 Moreover, we also obtain:
 \begin{align}\label{rhom}
 \rho_m (a) &= \frac{3\, H_I^2}{\kappa^2} \, \frac{(1-\nu)^2}{\alpha} \, \frac{D\, a^{3\, (1-\nu)\, (1 + \omega_m)}}{\Big(D\, a^{3\, (1-\nu)\, (1 + \omega_m)} + 1\Big)^2}, \nonumber \\
 \rho^{\rm RVM}(a) &=  \frac{3\, H_I^2}{\kappa^2} \, \frac{(1-\nu)}{\alpha} \, \frac{\nu D\, a^{3\, (1-\nu)\, (1 + \omega_m)}+1}{\Big(D\, a^{3\, (1-\nu)\, (1 + \omega_m)} + 1\Big)^2}.
 \end{align} 
 
>From \eqref{evolHsol} we observe that for the early-expanding-universe epoch,
where the fields are practically massless
(due to the high temperatures that we assume characterise the early RVM eras), one has $w_m \simeq 1/3$ and  
\begin{align}\label{early}
D \, a^{4(1-\nu)} \ll 1, 
\end{align}
and hence 
one obtains an approximate de-Sitter
solution\footnote{The reader's attention is called at this point to the fact that, as the scale factor $a(t) \to 0$ (Big Bang) in \eqref{evolHsol}, the RVM universe, whose evolution is based on Einstein's equations, does not exhibit an initial singularity~\cite{lima,perico,baslima,baslima2}. As we shall discuss later on in section \ref{sec:originGW}, this feature is also shared by our stringy RVM~\cite{ms1,ms1b}, but there, such a feature is due to stringy effects, such as the (infinity of the) higher-curvature contributions to the string-inspired effective gravitational action, which become important near the Big-Bang.}
\begin{align}\label{dS}
H(a)^{\rm dS} \simeq \Big(\frac{1 - \nu}{\alpha}\Big)^{1/2} \, H_I.
\end{align}
To ensure $H(a)^{\rm dS} \sim H_I $
during inflation, we may postulate
$\Big((1-\nu)/\alpha\Big)^{1/2} \simeq \mathcal O(1)$. As we shall discuss below, modern era phenomenology requires $\nu \ll 1$, which implies that 
\begin{align}\label{alpha}
\alpha^{1/2} \sim \mathcal O(1).
\end{align}

In such an early stage \eqref{early}, we observe from the first of Eqs.~\eqref{rhom} that the matter  density is almost vanishing, whilst the RVM energy density is approximately constant,
\begin{align}\label{matterearly}
\rho^{\rm early}_m \simeq 0, \quad \rho^{\rm RVM\, early}(a) \simeq  \frac{3\, H_I^2}{\kappa^2} \, \frac{(1-\nu)}{\alpha} \,,
\end{align}
as expected in a de Sitter phase.

As cosmic time progresses, $a$ becomes larger and larger, and the matter/radfiation densities started becoming non trivial, as a result of the {\it decay} of the running vacuum~\cite{lima,baslima,baslima2}.

Hence, the radiation (i.e. relativistic matter)-dominance era, which succeeds the inflationary era, is, characterised by $w_m=1/3$ but $D \, a^{4(1-\nu)} \gg 1$, and thus from 
\eqref{evolHsol} we have approximately:
\begin{align}\label{rad}
H(a)^{\rm rad} \simeq \Big(\frac{1 - \nu}{\alpha}\Big)^{1/2} \, \frac{H_I}{\sqrt{D} \, 
a^{2(1-\nu)}} \sim a^{-2},
\end{align}
as expected, since $\nu \ll 1$ and $\alpha^{1/2} = \mathcal O(1)$. 

Finally, in the modern era, which is dominated by a late dark energy contribution, during which the universe accelerates again, entering  another de-Sitter-type phase, we have that
$w_m \simeq 0$, because the matter is mostly non-relativistic. The  $H^4$  in \eqref{rgeh4} is not 
dominant, which, on account of \eqref{evolH}, implies:
\begin{align}\label{modern}
H^{\rm modern} (a) = H_0 \Big(  \Big[ 1 - \frac{c_0}{H_0^2\, (1-\nu)}  \Big] \, a^{-3\, (1-\nu)}  + \frac{c_0}{H_0^2\, (1-\nu)} \Big)^{1/2} \equiv H_0 \Big( \widetilde \Omega_{m\, 0} \, a^{-3(1-\nu)} + \widetilde \Omega_{\Lambda\, 0}\Big)^{1/2},
\end{align} 
where the quantities $\widetilde \Omega_{\Lambda\ 0} \equiv \frac{c_0}{H_0^2\, (1-\nu)} >0$ and 
$\widetilde \Omega_{m0} \equiv  1 - \widetilde \Omega_{\Lambda\, 0}$ play the r\^ole of the matter and cosmological-constant energy densities today, in units of the 
critical density of the universe, with the term $\widetilde \Omega_{\Lambda\, 0}$ {\it dominant} in the {\it current era}. The expression \eqref{modern} leads to observable deviations from  $\Lambda$CDM~\cite{sola6}.  Matching \eqref{modern} in the modern era with the relevant phenomenological data~\cite{sola5,sola6,tsiapi1,tsiapi2,sola11} yields 
\begin{align}\label{nucurr}
\nu \simeq {\mathcal O}(10^{-3}) \ll 1, 
\end{align}
as already announced previously. 

It should be stressed that, as becomes evident from the above analysis \eqref{evolHsol}, in the RVM framework it is essential that the coefficient $\nu \ne 1$. This is confirmed from the modern-era determination of $\nu$ \eqref{nucurr}, upon the assumption that the coefficients $\alpha, \nu$ remain constant during the entire evolution, which is the typical assumption within the conventional RVM framework. However, in the context of our string-inspired model~\cite{sola4,ms1,ms1b}, these coefficients are universe-era dependent, as there are appropriate phase transitions that take place between  the various epochs. Nonetheless the property that $1 - \nu \ne 0$ continues to hold in each era for the string-inspired RVM-like cosmological model as well. Indeed, as discussed in \cite{sola4}, and shall be reviewed below, during the inflationary era one has $\nu_{\rm infl} < 0,  \, |\nu_{\rm infl} | \ll 1$, and of course during the modern era $0 < \nu_0 \ll 1$. Hence, the essential features of the RVM evolution  are valid in its string-inspired extension, which we now proceed to discuss in some detail.

\section{Review of the string-inspired gravitational effective theory \label{sec:string}}

In string theory~\cite{string,string2}, the {\it massless} bosonic multiplet of the closed-string sector contains the spin-2 graviton field, described by the symmetric tensor $g_{\mu\nu}(x)=g_{\nu\mu}(x)$, the scalar (spin-0) dilaton $\Phi(x)$ and the spin-1 antisymmetric tensor (or Kalb-Ramond (KR)) field, $B_{\mu\nu}(x)=-B_{\nu\mu}(x)$. In the closed string sector, there is a U(1) gauge symmetry 
 \begin{align}\label{bu2}
B_{\mu\nu} \, \to \, B_{\mu\nu} + \partial_\mu \Theta_\nu (X) - \partial_\nu \Theta_\mu (X), 
\quad \mu, \nu =0, \dots 3, 
 \end{align}
where $\Theta_\mu(X)$, $\mu=0, \dots 3,$ are  gauge parameters, which 
implies that the target-space effective action, which describes the low-energy dynamics of the closed-string sector is invariant under \eqref{bu2}, and, as such, it depends only on the field strength 
of $B_{\mu\nu}$ :  $H_{\mu\nu\rho} = \partial_{[\mu} \, B_{\nu\rho]}$, where the symbol 
$[\dots ]$ indicates antisymmetrisation of the respective world indices. 

\subsection{String-inspired effective gravitational  action with torsion \label{sec:effstring}}

However the requirement of the cancellation between gauge and gravitational anomalies in the higher-dimensional spacetime of strings is achieved by 
the introduction of appropriate Green-Schwarz counterterms in the effective action~\cite{string,string2}, which results in the modification of the field strength $H_{\mu\nu\rho}$  by the Chern-Simons (gravitational (``Lorentz'', L)  and gauge (Y)) anomalous terms (in form language, for notational convenience):
\begin{align}\label{GSH}
\mathbf{{\mathcal H}} &= \mathbf{d} \mathbf{B} + \frac{\alpha^\prime}{8\, \kappa} \, \Big(\Omega_{\rm 3L} - \Omega_{\rm 3Y}\Big),  \nonumber \\
\Omega_{\rm 3L} &= \omega^a_{\,\,c} \wedge \mathbf{d} \omega^c_{\,\,a}
+ \frac{2}{3}  \omega^a_{\,\,c} \wedge  \omega^c_{\,\,d} \wedge \omega^d_{\,\,a},
\quad \Omega_{\rm 3Y} = \mathbf{A} \wedge  \mathbf{d} \mathbf{A} + \mathbf{A} \wedge \mathbf{A} \wedge \mathbf{A},
\end{align}
where $\wedge$ denotes the exterior product among differential ($k,\ell$) forms (${\mathbf f}^{(k)} \wedge {\mathbf g}^{(\ell)} = (-1)^{k\, \ell}\, {\mathbf g}^{(\ell)} \wedge {\mathbf f}^{(k)}$). In the above expression, $\mathbf{A}$ denote the Yang-Mills gauge field one form, and $\omega^a_{\,\,b}$ is the spin connection one form, with the Latin indices $a,b,c,d$ being tangent space (SO(1,3)) indices. The parameter $\alpha^\prime$ is the Regge slope, $\alpha^\prime = M_s^{-2}$ (in units of $\hbar=c=1$), with $M_s$ the string mass scale, which is in general different from the four-dimensional Planck scale, $M_s \ne M_{\rm Pl}$, and in fact it appears to be a free parameter of string theory, to be constrained phenomenologically.

The target-space-time low-energy string-inspire effective action is expanded in powers of $\alpha^\prime$. To lowest non-trivial order in $\alpha^\prime$, the (3+1)-dimensional effective action based on the aforementioned massless bosonic string multiplet, after appropriate compactification of the extra $n$-dimensions, reads~\cite{string,string2}:
 \begin{align}\label{sea}
S_B  =\; \int d^{4}x\sqrt{-g}\Big( \dfrac{1}{2\kappa^{2}} [-R + 2\, \partial_{\mu}\Phi\, \partial^{\mu}\Phi] - \frac{1}{6}\, e^{-4\Phi}\, {\mathcal H}_{\lambda\mu\nu}{\mathcal H}^{\lambda\mu\nu} + \dots \Big),
\end{align}
where we follow the conventions: $(+, -, -, -)$ for the metric signature, 
\begin{align}\label{riemann}
R^\lambda_{\,\,\mu\nu\sigma} = \partial_\nu \Gamma^\lambda_{\, \,\mu\sigma} + \Gamma^\rho_{\,\, \mu\sigma} \, \Gamma^\lambda_{\, \,\rho\nu} - (\nu \leftrightarrow \sigma)~,
\quad \lambda,\mu,\nu,\sigma= 0, \dots 3~,
\end{align} 
for the Riemann tensor,  $R_{\mu\nu} = R^\lambda_{\, \,\mu\lambda\nu}$ for the Ricci tensor, and $R=g^{\mu\nu} \, R_{\mu\nu}$ for the Ricci scalar. In \eqref{riemann} $\Gamma^\lambda_{\, \,\mu\sigma} =
\Gamma^\lambda_{\, \,\sigma\mu}$ is the Christoffel connection is the Riemannian one, in the absence of torsion, symmetric in its lower indices. The $\dots$ in 
\eqref{sea} denote terms of higher-orders in $\alpha^\prime$, which contain higher powers of curvature tensors and of (gravitational covariant) derivatives.\footnote{Throughout this review we shall concentrate on the lowest non-trivial order of string-effective actions, at most quadratic in derivatives acting on fields. In this limit, the torsion is non-propagating, as in Einstein-Cartan theory, discussed in the Appendix. However, once higher-derivative terms are taken into account, {\it e.g.} when $\mathcal O(\alpha^\prime)$ corrections are considered in the effective action~\cite{string,kaloper,amb1,amb2,amb3}, then derivative terms of the torsion $\mathcal H$ appear, including kinetic-like terms of $\mathcal H$. Taking into account such terms may lead to interesting phenomenology, which, however, will not affect the considerations in this review. We mention at this point that, in the (non-stringy) case of Einstein-Cartan theories, such considerations of kinetic terms of the totally antisymmetric part of the torsion have taken place in Ref.~\cite{benisty3}, with consequences for dark energy (from the kinetic term of the torsion per se), as well as for the existence of a bouncing cosmology solution, due to a stiff fluid that arises from the  quadratic (``mass-like'' terms) of the totally antisymmetric torsion dual $S_\mu S^\mu$ that characterise such theories, see discussion in our Appendix for the terminology (however in our contorted QED case we consider torsion induced by fermions~\cite{kaloper}, in contrast to the purely bosonic case of \cite{benisty3}. Moreover, in the stringy RVM model~\cite{sola4,ms1,ms2} it is the anomaly condensate that lead to inflationary-type dark energy, while higher-derivative terms  of the KR torsion $\mathcal H$ are suppressed for the scales relevant to the inflationary epoch in our case).} There is no bare cosmological constant in the effective action \eqref{sea} for strings living (before compactification) in their critical space-time dimension (in non-critical strings~\cite{aben,emn}, on the other hand, one may have such terms in case the dilatons are constant, otherwise one obtains relaxation-dark-energy (quintessence type) terms, with the dilaton as the quintenseence field). In our approach, such a cosmological constant term will arise dynamically through condensation of GW, as we shall discuss in the next section \ref{sec:gw}. 

In the string-inspired cosmological model of \cite{sola4,ms1,ms1b} the dilaton is self-consistently~\cite{ms2} assumed constant, $\Phi= \Phi_0$, $\partial_\mu \Phi =0$, and without loss of generality we can set its value to zero $\Phi_0 =0$. It is also assumed that 
in the early universe {\it only} fields from the {\it gravitational massless string multiplet} appear as {\it external fields}. This implies that we may consider the gauge fields $\mathbf A$ as absent in the early stages of the Universe evolution. 
Gauge fields, along with other chiral matter are assumed to be generated at the exit of inflation, as a consequence of the decay of the running vacuum, as we shall discuss later. 
So from now on we set $\mathbf A=0$ in the modification \eqref{GSH}. Then, one can arrive at the following Bianchi identity:
\begin{align}\label{modbianchi2}
 \varepsilon_{abc}^{\;\;\;\;\;\;\;\;\;\mu}\, \partial_\mu \, {\mathcal H}^{abc} 
 &=  \frac{\alpha^\prime}{32\, \kappa} \, \sqrt{-g}\, R_{\mu\nu\rho\sigma}\, \widetilde R^{\mu\nu\rho\sigma}  \equiv \sqrt{-g}\, {\mathcal G}(\omega) 
=  \frac{\alpha^\prime}{32\, \kappa} \, \sqrt{-g} \, {\mathcal K}^\mu (\omega)_{;\mu} =  \frac{\alpha^\prime}{32\, \kappa} \, \partial_\mu \Big(\sqrt{-g} \, {\mathcal K}^\mu (\omega) \Big) \nonumber \\
&  = \frac{\alpha^\prime}{16\, \kappa} \, \partial_\mu \Big[\epsilon^{\mu\nu\alpha\beta}\, \omega_\nu^{ab}\, \Big(\partial_\alpha \, \omega_{\beta ab} + \frac{2}{3}\, \omega_{\alpha a}^{\,\,\,\,\,\,\,c}\, \omega_{\beta cb}\Big)\Big],
\end{align}
where the semicolon ($;$) denotes gravitational covariant derivative with respect to the standard Christoffel connection without torsion,\footnote{In view of the total antisymmetry of $\mathcal H_{\mu\nu\rho}$, the gravitational covariant derivative with respect to the standard Christoffel symbol without torsion, symmetric in its lower indices,  acting on $\mathcal H_{\mu\nu\rho}$ coincides with the ordinary derivative, and this is indicated on the left-hand side of \eqref{modbianchi2}, where the ordinary derivative is given.} and
$ \varepsilon_{\mu\nu\rho\sigma} = \sqrt{-g}\,  \epsilon_{\mu\nu\rho\sigma}, \quad \varepsilon^{\mu\nu\rho\sigma} = \varepsilon_{\alpha\beta\gamma\delta} \,g^{\mu\alpha}\,g^{\nu\beta} \,g^{\rho\gamma} \,g^{\sigma\delta} =  
\frac{{\rm sgn}(g)}{\sqrt{-g}}\,  \epsilon^{\mu\nu\rho\sigma}$ (with Greek indices denoting space-time indices,
taking values $0, \dots, 3$) are the gravitationally covariant Levi-Civita tensor densities, totally antisymmetric in their indices, with $\epsilon_{\mu\nu\rho\sigma}$ ($\epsilon_{0123} = +1$, {\emph etc.}) the Minkowski-space-time Levi-Civita totally antisymmetric symbol. 
The symbol
$\widetilde{(\dots)}$
over the curvature tensor denotes the corresponding dual, $\widetilde R_{\mu\nu\rho\sigma} = \frac{1}{2} \varepsilon_{\mu\nu\alpha\beta} R_{\,\,\,\,\,\,\,\rho\sigma}^{\alpha\beta}$. In \eqref{modbianchi2} we explicitly denoted the total-derivative character of the gravitational-anomaly Chern-Simons terms $\sqrt{-g}\,{\mathcal G}(\omega)$~\cite{anom}. 

Before proceeding, we notice that  the quadratic $\mathcal H$-terms in \eqref{sea} can be absorbed in a generalised curvature scheme with torsion (see Appendix) 
\begin{align}\label{actiontorsion}
S_B = - \frac{1}{2\kappa^2} \, \int d^4 x \sqrt{-g} \, \widehat R(\overline \Gamma) + \dots~,
\end{align} 
where the generalised Ricci scalar is defined as 
$\widehat R(\widehat \Gamma) = g^{\mu\sigma} \delta^\nu_{\, \,\lambda} \, \widehat R^\lambda_{\,\, \mu\nu\sigma}(\widehat \Gamma)$, with the generalised curvature Riemann tensor $\widehat R^\lambda_{\, \,\mu\nu\sigma}(\widehat \Gamma)$ defined as in \eqref{riemann} but with 
the ordinary (symmetric in its lower indices) Christoffel symbol replaced by the 
torsional connection
\begin{align}\label{torsion}
{\overline \Gamma}_{\mu\nu}^{\rho} = \Gamma_{\mu\nu}^\rho + \frac{\kappa}{\sqrt{3}}\, {\mathcal H}_{\mu\nu}^\rho  \ne {\overline \Gamma}_{\nu\mu}^{\rho}
\end{align}
with $\Gamma_{\mu\nu}^\rho = \Gamma_{\nu\mu}^\rho$ the torsion-free Christoffel symbol. Since the KR field strength satisfies $\mathcal H^\mu_{\nu\rho} = -
\mathcal H^\mu_{\rho\nu}$, it plays the r\^ole of the contorsion~\cite{torsion}. 
This contorted geometry 
contains only a totally antisymmetric component of torsion.\footnote{Using local field redefinition ambiguities~\cite{string,string2,kaloper,amb1,amb2,amb3}, which leave the perturbative string scattering amplitudes unaffected, according to the equivalence theorem~\cite{equiv1,equiv2}, one can extend the torsion interpretation of $\mathcal H$ to 
${\mathcal O}(\alpha^\prime)$ effective actions, which contain fourth-order derivative terms.} This is a distinguishing feature of the string model from other generic torsion cosmologies (e.g. \cite{tors1,tors2}), in which the torsion has more components.
Moreover, since the string multiplet necessarily contains a graviton field, this string-inspired gravitational theory \eqref{sea} is different from teleparallel gravity~\cite{tele,heis1,heis2,heis3,tele2}, where torsion mimics the gravitational field.

\subsection{Connection with torsional topological invariants and axions \label{sec:torsinv}}

An important comment is now in  order concerning the form of the string-inspired contorted action \eqref{actiontorsion}. The reader observes that there is no Holst term
\begin{align}\label{holst} 
S_{\rm Holst} \propto  \int d^4x \, \varepsilon^{\mu\nu\rho\sigma} \, \widehat R_{\mu\nu\rho\sigma} (\overline \Gamma), 
\end{align}
which, in other approaches to QG, like LQG~\cite{loop,loop2} and spin-foam models~\cite{foam,foam2}, as mentioned in the 
introduction, carries as a coefficient 
the Barbero-Immirzi parameter~\cite{bi,bi2}. The absence of such a term from the perturbative string-amplitude approach that leads to \eqref{actiontorsion} can perhaps be interpreted as 
implying that the string effective action is linked with the 
so-called Nieh-Yan invariant~\cite{nieh,nieh2,nieh3,nieh4,nieh5} rather than the Holst term,\footnote{In fact, as emphasised in \cite{mercuri,mercuri2}, the Holst term alone is not a topological invariant, unlike the Nieh-Yan term, which thus seems more appropriate to use in order to bring into (a non-perturbative) play the Barbero-Immirzi parameter in continuous (contorted) gravitational effective actions, stemming from microscopic QG models.} which is a total derivative even in  the presence of torsion and replaces the Holst term~\cite{mercuri,mercuri2} 
\begin{align}\label{nieh}
S_{\rm Nieh-Yan} \propto \int d^4 x  \, \Big(\varepsilon^{\mu\nu\rho\sigma} \, T^\lambda_{\,\,\,\,\mu\rho} \, T_{\lambda\nu\sigma} - \varepsilon^{\mu\nu\rho\sigma} \, \widehat R_{\mu\nu\rho\sigma} (\overline \Gamma) \Big)=  \int d^4 x  \, \partial_\mu \Big(\varepsilon^{\mu\nu\rho\sigma} \, T_{\nu\rho\sigma} \Big),  
\end{align}
where $T^\mu_{\,\,\,\,\nu\rho}$ is the torsion tensor (see Appendix), which in the context of the Einstein-Cartan theory is a non-propagating field, as the corresponding action contains non-derivative terms of the torsion tensor (the graviton of course is a propagating massless spin-2 field, independent of the torsion in this formalism). Such terms, therefore cannot contribute to the perturbative string-scattering amplitudes, but may exist in non-perturbative formulations of string theory.  In our case, the torsion, as already mentioned, has a single totally antisymmetric component, proportional to the KR field strength $T_{\mu\nu\rho} \propto \mathcal H_{\mu\nu\rho}$, and thus the Nieh-Yan invariant is nothing other but the left-hand-side of the Bianchi identity \eqref{modbianchi2}, that is, the gravitational Chern Simons term. 

We shall argue next that the totally antisymmetric torsion corresponds to a massless pseudoscalar degree of freedom, which in the string context is the so-called string-model independent or KR axion field~\cite{kaloper,svrcek}. We note that the association of an axion with the totally antisymmetric component of a torsion is a generic feature of contorted models, as discussed in the Appendix, where the example of contorted QED is studied. 
To this end, we implement the Bianchi identity as a $\delta$-functional constraint 
in the quantum path-integral of the action \eqref{sea}, where the partition function is expressed as a path integral over the graviton and $\mathcal H$ fields: 
\begin{align}\label{pathintegral}
\mathcal Z &= \int {\mathcal D} g {\mathcal D} {\mathcal H} \, 
\delta \Big( \varepsilon_{abc}^{\;\;\;\;\;\;\mu}\, \partial_\mu \, {\mathcal H}^{abc}
 -  \frac{\alpha^\prime}{32\, \kappa} \, \sqrt{-g}\, R_{\mu\nu\rho\sigma}\, \widetilde R^{\mu\nu\rho\sigma} \Big)\, 
\exp \Big(i\, \int d^4 x \sqrt{-g} \Big[-\frac{1}{2\kappa^2} R - \frac{1}{6}\,  {\mathcal H}_{\lambda\mu\nu}
{\mathcal H}^{\lambda\mu\nu} \Big)\Big] \nonumber \\
&= \int {\mathcal D} g {\mathcal D} {\mathcal H} \, {\mathcal D} b \, 
\exp \Big(i\, \int d^4 x \sqrt{-g} \Big[-\frac{1}{2\kappa^2} R - \frac{1}{6}\,  {\mathcal H}_{\lambda\mu\nu}
{\mathcal H}^{\lambda\mu\nu}  - (\partial_\mu  b)\, \epsilon_{abc}^{\;\;\;\;\;\;\mu}\, {\mathcal H}^{abc} -  b\, \frac{\alpha^\prime}{32\, \kappa} \, R_{\mu\nu\rho\sigma}\, \widetilde R^{\mu\nu\rho\sigma} \Big]  \Big),
\end{align} 
where the $b(x)$  in the second line is a (canonically normalised) pseudoscalar (axion-like) Lagrange-multiplier field, which is implementing the delta-functional constraint in the path-integral, and we have performed appropriate integrations by parts in the exponent. After the quadratic 
$\mathcal H$ path integration, we end up easily with an effective theory of a fully dynamical axion field (KR or string-model independent axion) $b(x)$, in a Riemannian curved space-time, with  effective action
\begin{align}\label{sea4}
S^{\rm eff}_B &=
\; \int d^{4}x\, \sqrt{-g}\Big[ -\dfrac{1}{2\kappa^{2}}\, R + \frac{1}{2}\, \partial_\mu b \, \partial^\mu b  +  \sqrt{\frac{2}{3}}\,
\frac{\alpha^\prime}{96 \, \kappa} \, b(x) \, R_{\mu\nu\rho\sigma}\, \widetilde R^{\mu\nu\rho\sigma}
   + \dots \Big]
\nonumber \\
& = \; \int d^{4}x\, \sqrt{-g}\Big[ -\dfrac{1}{2\kappa^{2}}\, R + \frac{1}{2}\, \partial_\mu b \, \partial^\mu b  -
 \sqrt{\frac{2}{3}}\,
\frac{\alpha^\prime}{96 \, \kappa} \, {\mathcal K}^\mu (\omega)\, \partial_\mu b(x)   + \dots \Big],
\end{align}
where $\mathcal K^\mu (\omega)$ has been defined in \eqref{modbianchi2}, and expresses the total derivative of the gravitational Chern-Simons anomalous terms. 
It is noted that the pseudoscalar nature of the Lagrange multiplier field $b(x)$ is necessitated by the fact that the Chern-Simons terms violate CP symmetry, and thus the CP invariance of the gravitational Lagrangian \eqref{sea4} (and \eqref{sea}) is guaranteed only if $b$ is pseudoscalar (the $\delta$-functional constraint of \eqref{modbianchi2} is by construction CP invariant).

A few important comments are now in  order, concerning the effective action \eqref{sea4}. In \cite{castel,taveras,taveras2,taveras3}, it was suggested to promote the BI parameter, which is assumed to be the coefficient of the Nieh-Yan  invariant, to a dynamical pseudoscalar field. In view of 
\eqref{modbianchi2}, in our string case, this will lead to 
the reduction of the Nieh-Yan invariant to the gravitational anomaly term, and thus to a coupling of this BI axion field to the gravitational Chern-Simons term. In fact, 
as we shall now come to discuss, the BI axion field is essentially the string-model independent axion field~\cite{kaloper,svrcek}, upon taking proper account, however,  of the Green-Schwarz counterterms appearing in \eqref{modbianchi2}. 

Indeed, following \cite{taveras,taveras2,taveras3}, promotion of the BI parameter to a field would naively be  equivalent to considering adding to the effective action \eqref{sea} a term \eqref{nieh} but with a coordinate-dependent coefficient $\beta (x)$, which is viewed as a field variable, integrated over in the path-integral (with a measure $\mathcal D \beta(x)$):
\begin{align}\label{nieh2}
S_{\rm Nieh-Yan}^{\rm BI-field} &=  \int d^4 x\,  \beta(x) \, \Big(\varepsilon^{\mu\nu\rho\sigma} \, T^\lambda_{\,\,\,\,\mu\rho} \, T_{\lambda\nu\sigma} - \varepsilon^{\mu\nu\rho\sigma} \, \widehat R_{\mu\nu\rho\sigma} (\overline \Gamma) \Big)
\nonumber \\
&=  \int d^4 x  \, \beta (x)\, \partial_\mu \Big(\varepsilon^{\mu\nu\rho\sigma} \, T_{\nu\rho\sigma} \Big) \propto  \int d^4 x  \, \beta (x) \,  \partial_\mu \Big(\varepsilon^{\mu\nu\rho\sigma} \, \mathcal H_{\nu\rho\sigma} \Big)
\end{align}
where in the last equality the proportionality factor takes into account the precise relation between the torsion and the KR field strength \eqref{torsion}. Integrating over (the non-propagating field) $\mathcal H$ in the effective action \eqref{sea}, then, would produce a dynamical propagating field $\beta (x)$, with kinetic terms canonically normalised, which is the result of \cite{taveras,taveras2,taveras3}. However, the above procedure would imply a Bianchi identity constraint $\varepsilon^{\mu\nu\rho\sigma} \, \partial_\mu\,\mathcal H_{\nu\rho\sigma} =0$, which does not take into account 
the Green-Schwarz counteterms \eqref{modbianchi2}. Implementing the correct constraint \eqref{modbianchi2}, as we did above ({\it cf.} \eqref{pathintegral}), leads to the correct formulation of the association of the totally antisymmetric torsion with a dynamical axion-like field. 

Thus, in the string case, the promotion of the BI parameter accompanying the Nieh-Yan invariant \eqref{nieh} into a dynamical pseudoscalar field cannot be done solely via \eqref{nieh2}, as in \cite{taveras,taveras2,taveras3}, but by considering instead adding to the effective action \eqref{sea} an {\it appropriate combination} of topological invariants, the Nieh-Yan invariant and the gravitational Chern-Simons term, with the BI field appearing as its coefficient. That is, we   
add to the effective action \eqref{sea} the combination of topological invariants:
\begin{align}\label{nieh3}
S_{\rm Nieh-Yan}^{\rm BI-field} + S_{\rm Grav.~Chern~Simons} &=  
 \int d^4 x  \, b(x) \, \Big(\varepsilon^{\mu\nu\rho\sigma} \, \partial_\mu \, \mathcal H_{\nu\rho\sigma}  -  \frac{\alpha^\prime}{32\, \kappa} \, \sqrt{-g}\, R_{\mu\nu\rho\sigma}\, \widetilde R^{\mu\nu\rho\sigma} \Big),
\end{align}
generalising appropriately the construction of \cite{taveras,taveras2,taveras3},  
where now $b(x)$ is the Lagrange multiplier  implementing  the Green-Schwarz-modified Bianchi constraint \eqref{modbianchi2} in the path-integral~\cite{kaloper,svrcek}, which 
becomes the string-model independent axion, as we discussed above ({\it cf.} \eqref{pathintegral}). We remark for completeness, that, as an axion, the field $b(x)$ respects the shift symmetry $b(x) + {\rm constant}$, which is a featured shared by the BI field in the approach of \cite{taveras2}. Further discussion on this point can be found in the Appendix. Eq \eqref{nieh3} contains therefore two propagating independent massless degrees of freedom, the KR axion $b(x)$ and the graviton. 

We therefore conjecture, at this stage, that the promotion of the BI parameter as an axion field, 
via \eqref{nieh3}, makes the low-energy string-effective action \eqref{sea4},  based on degrees of freedom in the massless gravitational multiplet of strings, probably obtainable also as the low-energy continuous limit of other QG approaches, such as (appropriate extensions/modifications of) LQG and spin-foam models, once torsion and gravitational anomalies are properly accounted for. We next note that, on  taking into account the SL(2,Z) symmetry of strings~\cite{string,string2}, the axion field will necessarily be accompanied also by a dilaton field in the effective action, which can be accommodated in such alternative approaches to QG as well (but, as we already discussed, the dilaton can also be set self-consistently to a constant~\cite{ms2,ms3}).

\subsection{Gravitational anomalies, axions and the r\^ole of the Cotton tensor \label{sec:anom}}

Thus, the effective action \eqref{sea4} contains complete information on the dynamics of the KR torsion in the context of string theory, and this is the action we shall concentrate upon from now on, to discuss the association of the pertinent cosmology with the running vacuum, following \cite{sola4,ms1,ms1b}.\footnote{It should be noted at this point that in string theory there are several other types of axions, arising from compactification~\cite{svrcek}, which lead to a rich phenomenology~\cite{arvanitaki,marsh}. We discussed such issues in \cite{ms1,ms1b}, but we shall not describe them here. These additional stringy axions will not affect the basic objective of our approach~\cite{sola4}, which is to demonstrate the conditions under which the string-inspired cosmological model \eqref{sea4}  
reduces to a RVM cosmology~\cite{sola,sola2,sola3}.}
To this end, we first observe that the gravitational variation of the Chern-Simons term is non trivial, giving rise to the Cotton tensor $\mathcal C_{\mu\nu}$~\cite{jackiw}:\footnote{It should be noted, for completeness, that this non-trivial contribution of the gravitational Chern-Simons term to the stress tensor of the axion matter does not characterise the axion coupling with gauge Chern-Simons terms 
$\int d^4x\, \sqrt{-g}\, b \, {\rm Tr} (\mathbf F_{\mu\nu} \widetilde{\mathbf F}^{\mu\nu})$ 
(with the Tr referring to gauge-group indices), whose gravitational variation vanishes.
}   
\begin{align}
\delta \Big[ \int d^4x \sqrt{-g} \, b \, R_{\mu\nu\rho\sigma}\, \widetilde R^{\mu\nu\rho\sigma} \Big] = 4 \int d^4x \sqrt{-g} \, {\mathcal C}^{\mu\nu}\, \delta g_{\mu\nu} = -
4 \int d^4x \sqrt{-g} \, {\mathcal C}_{\mu\nu}\, \delta g^{\mu\nu}, 
\end{align}
where 
\begin{align}\label{cotton}
{\mathcal C}^{\mu\nu} \equiv  -\frac{1}{2}\, \Big[v_\sigma \, \Big( \varepsilon^{\sigma\mu\alpha\beta} R^\nu_{\, \, \beta;\alpha} +
\varepsilon^{\sigma\nu\alpha\beta} R^\mu_{\, \, \beta;\alpha}\Big) + v_{\sigma\tau} \, \Big(\widetilde R^{\tau\mu\sigma\nu} +
\widetilde R^{\tau\nu\sigma\mu} \Big)\Big], 
\end{align}
which is tracelss: 
\begin{align}\label{trace}
g^{\mu\nu}\, \mathcal C_{\mu\nu} =0,
\end{align}  
with $v_{\sigma} \equiv \partial_\sigma b = b_{;\sigma}, \,\,v_{\sigma\tau} \equiv  v_{\tau; \sigma} = b_{;\tau;\sigma}$. Thus, Einstein's equations stemming from \eqref{sea4} read
\begin{align}\label{einsteincs}
R^{\mu\nu} - \frac{1}{2}\, g^{\mu\nu} \, R  -  \sqrt{\frac{2}{3}}\,
\frac{\alpha^\prime \, \kappa}{12} \,  {\mathcal C}^{\mu\nu}  = \kappa^2 \,
\Big(\partial^\mu b \, \partial^\nu b - \frac{1}{2} \, g^{\mu\nu} \, \partial_\alpha b \, \partial^\alpha b  \Big) \equiv \kappa^2 \, T^{\mu\nu}_{b} ,
\end{align}
with the right-hand-side being the stress energy tensor of the KR axion, which plays the r\^ole of ``matter'' in this set up (of course, the $b$ the field being is associated with a massless excitation of the string gravitational multiplet in this case). Because of  the property of the Cotton tensor \eqref{cotton}~\cite{jackiw}
\begin{align}\label{propcotton}
\mathcal C^{\mu\nu}_{\,\,\,\,\,\,;\mu} = -\frac{1}{8} \, v^\nu \, R^{\alpha\beta\gamma\delta}\, 
\widetilde R_{\alpha\beta\gamma\delta}\,,
\end{align}
we obtain, from the Bianchi identity of the covariant derivative of the Einstein tensor, the following generalised conservation law of ``matter'' in this system:
\begin{align}\label{bcons}
T^{\mu\nu}_{b\, \,\,\,; \mu} +  \sqrt{\frac{2}{3}}\,
\frac{\alpha^\prime}{12 \, \kappa} \, {\mathcal C}^{\mu\nu}_{\,\,\,\,\,\,\,;\mu} =0,
\end{align}
which, implies that in the presence of non-trivial gravitational anomalies, there is an exchange of energy between ``matter'' (KR axion $b$ in this particular case) and gravity, through the gravitational anomalous interactions. This apparent generalised conservation law is perfectly consistent with the general covariance of the system~\cite{sola4,ms1,ms1b}, as is evidenced from the above equations, and in fact the exchange of energy depends on the kind of background space time one encounters. Indeed, for a FLRW universe, the anomalous terms and the Cotton tensor both vanish, and hence the right-hand-side of \eqref{propcotton} is zero, leading to the conventional conservation law of $b$ matter.
This is not the case, however, as we shall see below, if there are CP-violating GW perturbations in the FLRW space time, in which case the anomaly terms and the Cotton tensor are non zero. In such perturbed geometries, there will be a non-trivial exchange of energy between  the b field and the gravity sector. 

\section{Primordial Gravitational Waves, Anomalies and an RVM-like Inflation without inflatons \label{sec:gw}}

One of the basic ingredients that ensures the non-vanishing of the gravitational Chern Simons terms is the presence of primordial GW.  In \cite{ms1,ms1b}, we have discussed various scenarios which allow the presence of GW in the primordial string universe, and refer the interested reader there for details. In what follows, we shall only briefly sketch these scenarios so as to give complete information to the reader as to the potential origin of GW, which, as we shall discuss later on in this section, play a crucial r\^ole in the connection of the string-inspired cosmological model based on \eqref{sea4} to the RVM~\cite{sola4}.

\subsection{On the origin of primordial Gravitational Waves \label{sec:originGW}} 

One of the most obvious reasons for obtaining GW is the non-spherical merging of rotating primordial black holes, that may characterise the very early string Universe. Indeed, such rotating black holes with secondary axion charge do arise as a solution of the gravitational string-inspired theory \eqref{sea4}~\cite{kaloper}, where the rotating nature is linked to the axial character of the KR axion. It is therefore conceivable that such primordial black holes arise as a result of vacuum space-time fluctuations themselves in the context of the model of \cite{sola4}, where the early Universe dynamics contains no degrees of freedom 
other than the massless KR axion and gravitons. In more general scenarios, involving brane universes~\cite{ms1,ms1b}, punctured with appropriate massive space-time defects consisting of D-branes compactified along the extra spatial dimensions, which, from a low-energy brane-world observer point of view,  look ``effectively point-like"~\cite{westmuckett,westmuckett2,westmuckett3}, GW arise from coalescent rotating black holes which arise from the gravitational collapse of these massive defects.

\begin{figure}[!h]
%[!h]
\centering
\includegraphics[width=4.5in]{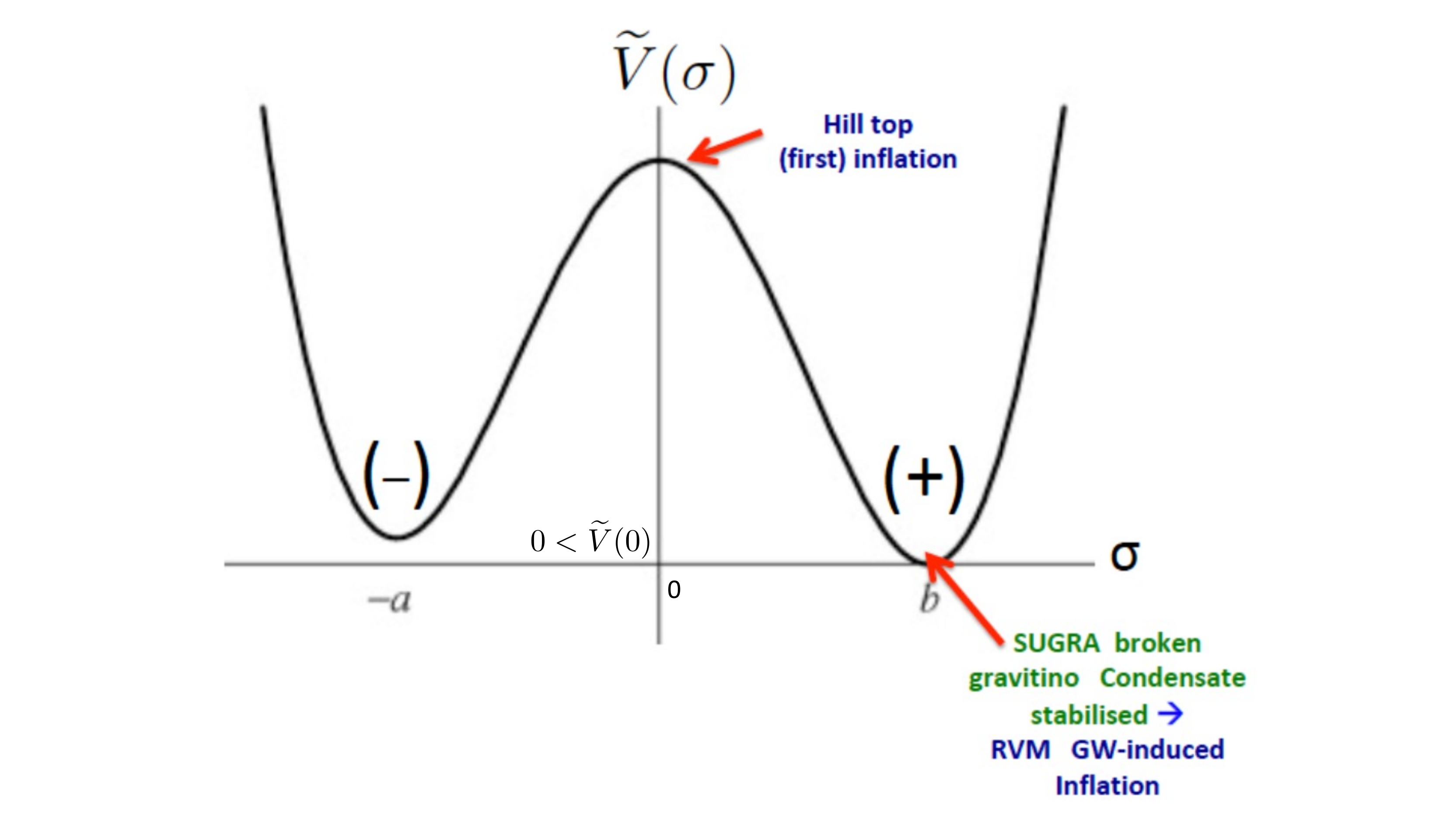}
%%%% where xxxxxx name represents "figurename.eps"
\caption{The double-well gravitino-condensate potential $\widetilde V(\sigma)$ vs. the condensate field $\sigma$, for dynamical supergravity breaking~\cite{houston,houston2,ms1,ms1b}, in which there is an effective lift of the degeneracy of the two vacua due to percolation effects in the early Universe that lead to different occupation probabilities for the two ground states~\cite{ovrut,ovrut2}. The reader should notice that the values of the potential at these ground states are both positive, consistent with the dynamical breaking of supergravity~\cite{houston,houston2}. 
This bias between the two vacua leads to the formation of unstable domain walls, whose non-spherical collapse, or collisions produce GW. The lowest of these two vacua indicates dynamical breaking of supergravity, with a stabilised gravitino condensate, which also implies GW-induced RVM inflation. The model may have a hill-top first inflation, near the origin of the gravitino condensate field $\sigma \simeq 0$, as indicated, which ensures that any spatial inhomogeneities are washed out well before the entrance of the Universe intro the RVM inflationary phase, induced by GW condensation. Figure taken from \cite{ms1,ms1b}.}
\label{fig:sugra}
\end{figure}

On the other hand, as emphasised in \cite{ms1,ms1b}, there are also simpler scenarios, which involve only a supergravity extension of the  gravitational theory \eqref{sea4}, which arises in superstrings. The gravitino, which is the spin-3/2 superpartner of gravitons, still belongs to the (initially) massless string gravitational multiplet, and hence such excitations are consistent with the point of view of \cite{sola4} that only gravitational degrees of freedom appear as external fields in the effective gravitational action describing the dynamics of the very early universe. One may then have, under certain conditions, the formation of gravitino condensates, $\sigma = \langle \overline \psi_\mu \, \psi^\mu \rangle $, 
which could break the supergravity dynamically, as in \cite{houston,houston2}.  The condensate field 
$\sigma (x)$ which describes quantum excitations about the condensate, has then a double-well potential, and may lead to unstable domain wall (DW) formation, if there is a slight lift of the degeneracy of the vacua, {\it e.g.} due to percolation effects in the early Universe, which could lead to a statistical bias induced by different occupation probabilities between the two ground states (see fig.~\ref{fig:sugra}). Non-spherical collapse or collision of such DW leads then to the formation of GW. 

This dynamically-broken supergravity scenario
has another desirable feature apart from 
leading to DW formation. It may be characterised by a hill-top inflation~\cite{hilltop}, not necessarily slow-roll (see fig.~\ref{fig:sugra}), and without phenomenological consequences, which serves the important purpose of washing out any spatial inhomogeneities of the KR axion and graviton fields, thus providing a physical justification of the isotropy and homogeneity of the string Universe well before its entrance to a second, slow-roll, RVM inflationary phase, induced by GW condensation.~\footnote{We remark for completion that the hill-top first inflation is assumed to occur soon after the Big-Bang. However, in the context of string theories, whose effective point-like low-energy gravitational field theories contain in principle an infinity of higher-derivative, higher-curvature terms, there may not be an initial singularity. This feature is already demonstrated at the level of dilaton-Gauss-Bonnet modified gravity theories~\cite{art}, which are embeddable in string theories. As we have already mentioned in section \ref{sec:rvm}, the absence of initial singularities also characterises the RVM Cosmology. Such issues will not be of concern to us in the current work, as we are interested in the epochs well after the Big-Bang, where the effective field theory \eqref{sea4} suffices for the description of the pertinent dynamics.} It is this second RVM inflation that has phenomenological consequences and can be constrained by cosmological data~\cite{Planck}. This justifies the assumptions and analysis of \cite{sola4}. 

\subsection{Gravitational-wave condensation, gravitational Chern-Simons terms and RVM inflation \label{sec:rvminfl}}

In the presence of GW perturbations, the Chern-Simons term is non trivial~\cite{stephon}.
In this subsection, we shall assume first that we are in an inflationary phase, where the Hubble parameter is approximately constant, and evaluate the corresponding condensate of the gravitational Chern-Simons term by integrating out graviton fluctuations up to a momentum cut-off scale $\mu$. Then we shall demonstrate~\cite{sola4}, based on generic properties of the Cotton tensor \eqref{cotton}, that the cosmological vacuum satisfies indeed a de Sitter equation of state, while the vacuum energy density  assumes an RVM-like form \eqref{rgeh4}, with a non-trivial dominant $H^4$ contribution, but a negative coefficient of the $H^2$ terms, due to non-positive contributions of the gravitational anomaly to the stress tensor of theory. Inflation then, without inflaton fields, which we assumed as a background, is justified {\it a posteriori} self-consistently by the early Universe solution of the evolution equation \eqref{evolH} for the Hubble parameter that characterises 
the RVM. 

Let us review the situation concretely below. We shall be brief, since, for details, we refer the interested reader in \cite{sola4,ms1,ms1b}.  On assuming GW perturbations on a FLRW  inflationary background space-time, with constant Hubble parameter $H_I$, is equivalent to considering the following metric in the FLRW frame:
\begin{align}\label{GWmetric}
 ds^2 = dt^2 - a^2(t) \Big[(1 - h_+(t,z))\, dx^2 + (1 + h_+(t,z))\, dy^2 + 2h_\times (t,z)\, dx\, dy + dz^2 \Big], \quad  a(t) \sim e^{H_I\, t},
 \end{align}
in standard notation for the polarizations of the GW, assumed propagating along the z-direction for concreteness. Integrating over GW perturbations (graviton modes), with spatial momenta of magnitude $k$  up to an Ultra-Violet (UV)  cutoff $\mu$, one obtains for  the gravitational-anomaly condensate~\cite{stephon,sola4}:
 \begin{align}\label{rrt2}
  \langle R_{\mu\nu\rho\sigma}\, \widetilde R^{\mu\nu\rho\sigma} \rangle  &\simeq  
  \frac{16}{a^4} \, \kappa^2\int^\mu \frac{d^3k}{(2\pi)^3} \, \frac{H_I^2}{2\, k^3} \, k^4 \, \Theta  = \frac{1}{\pi^2} \Big(\frac{H_I}{M_{\rm Pl}}\Big)^2 \, \mu^4\, \Theta   \nonumber \\
 &= \frac{2}{3\pi^2} \frac{1}{96 \times 12} \,  \Big(\frac{H_I}{M_{\rm Pl}}\Big)^3 \, \Big(\frac{\mu}{M_{\rm Pl}}\Big)^4 \,  M_{\rm Pl}\, \times \, \, {\mathcal K}^0 (t)\,,
\end{align}
to leading order in the slow-roll parameter 
\begin{align}\label{theta}
 \Theta = \sqrt{\frac{2}{3}}\, \frac{\alpha^\prime \, \kappa}{12} \, H_I \,  {\dot {\overline b}} \, \ll \, 1~,
  \end{align}
with the overdot denoting derivative with respect to the cosmic time $t$ in the RW frame. The covariant anomaly four vector $\mathcal K^\mu (t)$, assumed isotropic and homogeneous,  is defined in \eqref{modbianchi2}. We stress again that this isotropy and homogeneity can be justified by the aforesaid scenario of having a first hill-top inflation before the RVM-like one~\cite{ms1,ms1b}.

>From the anomaly equation \eqref{modbianchi2}, assuming homogeneity and isotropy, we 
have that 
\begin{align}\label{anomeq}
\frac{d}{dt} \mathcal 	K^0 (t)= \langle R_{\mu\nu\rho\sigma}\, \widetilde R^{\mu\nu\rho\sigma} \rangle, 
\end{align} 
from which, using \eqref{rrt2}, we  
easily  arrive at an evolution equation for $\mathcal K^0(t)$, which admits as a solution:
\begin{eqnarray}\label{k02}
{\mathcal K}^0 (t) = {\mathcal K}^0_{\rm begin} (t=0) \, \exp\Big[  - 3H_I\, t \, \Big( 1  -  
\frac{1}{3\,\pi^2 \times 18 \times  96}\, \Big(\frac{H_I}{M_{\rm Pl}}\Big)^2 \, \Big(\frac{\mu}{M_s}\Big)^4 \Big)\Big],
\end{eqnarray}
where ${\mathcal K}^0_{\rm begin} (t=0)$ is a boundary condition which can be determined phenomenologically in the context of low-energy string effective actions~\cite{sola4,ms1,ms1b}.
We observe from \eqref{k02} that an approximately constant $\mathcal K^0$ throughout the duration of inflation can be obtained under the condition of an approximately vanishing exponent on the right-hand-side of \eqref{k02}, which amounts to the approximate relation (to be understood as an order of magnitude relation):
\begin{align}\label{k0const}
\frac{\mu}{M_s} \simeq 15 \, \Big(\frac{M_{\rm Pl}}{H_I}\Big)^{1/2}.
\end{align}
The Planck-Collaboration-data result \eqref{PlHI} for the inflationary scale~\cite{Planck},  then, 
 combined with \eqref{k0const}, imply  
 \begin{align}\label{mMs}
 \mu \simeq \mathcal O(10^3)\,M_s. 
 \end{align}

Due to the homogeneity and isotropy of this early universe, the equation of motion for the $b$ field, stemming from \eqref{sea4}, admits the 
solution
\begin{align}\label{krbeom2}
\dot{\overline{b}}  =  \sqrt{\frac{2}{3}}\, \frac{\alpha^\prime}{96 \, \kappa} \, {\mathcal K}^{0},
\end{align}
which for approximately constant $\mathcal K^0$ implies $\dot b \simeq {\rm contant}$. This violates {\it spontaneously}  Lorentz invariance. Parametrising this solution  
as
\begin{align}\label{slowrollb} 
\dot b = \sqrt{2\epsilon} H_I  M_{\rm Pl},
\end{align}
with $\epsilon$ a phenomenological constant parameter, 
we observe~\cite{sola4} that \eqref{slowrollb} is 
consistent with the Planck Collaboration slow-roll data~\cite{Planck}, provided we set
\begin{align}\label{epsilonref} 
\epsilon = \mathcal O(10^{-2}).
\end{align}
This implies 
\begin{align}\label{slowrollbint}  
b(t) = \overline b(0) + \sqrt{2\epsilon} H_I  M_{\rm Pl} \, t, \quad H_I \simeq {\rm constant}, 
\end{align}
with $\overline b(0)$ the value of the field $b(x)$ at the onset of inflation at cosmic time $t=0$.

A few remarks are now in order. If we insist on the theory respecting the transplanckian conjecture, that is, that there are no modes in the theory with momenta higher than the Planck scale, which acts as the ultimate UV cutoff of the effective theory,\footnote{Assuming the transplanckian censorship hypothesis (TCC) in inflationary cosmology~\cite{tcc1}, {\it i.e.} that ``{\it any inflationary model which can stretch the quantum fluctuations with wavelengths smaller than the Planck length scale out of the Hubble horizon is in the swampland}'', then it was argued in \cite{tcc2} that this would imply very small Hubble scales during inflation, which would in turn lead to negligible amplitudes for primordial GW. However, one of the two basic assumptions of this conjecture is that there is instantaneous reheating of the Universe right after inflation. Relaxing the instantaneous reheating assumption, as happens in some low-reheating cosmological models, lead the authors of Ref.~\cite{avoidtcc} to bypass the aforementioned constraints, and obtain high allowed upper bounds on the inflationary scale, inversely proportional to the (low) reheating 
temperature (in units of today's CMB temperature), while respecting the TCC. In the RVM~\cite{sola,sola2,sola3,lima,perico,baslima,baslima2} and stringy-RVM~\cite{sola4,ms1,ms1b,ms2} frameworks, there is no reheating of the Unviverse during the RVM-vacuum decay that occurs at the exit of the RVM inflation. Thus, such assumptions can be avoided, and, thus, the scale of the stringy RVM inflation, of interest to us in this review, can be high, in agreement with the standard phenomenology ({\it cf.}\eqref{PlHI})~\cite{ms1}.} and combine this requirement with the slow-roll condition for the $b$-field \eqref{theta}, and \eqref{slowrollb}, \eqref{epsilonref},  we obtain that the  
 string mass scale is restricted to lie in the range~\cite{ms1,ms1b} 
 \begin{align}\label{msscale}
 2.6 \times 10^{-3}\, M_{\rm Pl}  \le M_s \gg10^{-5}\, M_{\rm Pl}~.
 \end{align} 
An average order of magnitude of $M_s$, then, which satisfies \eqref{msscale}, is 
\begin{align}\label{msmpl}
M_s = \mathcal O(10^{-3}) \, M_{\rm Pl}, 
\end{align} 
which we may assume for the rest of this article. On account of \eqref{mMs}, then, we may deduce that $\mu \sim M_{\rm Pl}$, that is, the UV cutoff of the effective theory is of order of the subplanckian string scale \eqref{msmpl}.

A direct consequence of \eqref{slowrollbint}  is that 
$\dot b$ remains undiluted at the end of inflation and thus non trivial well onto the radiation era~\cite{sola4,ms1,ms1b}. As we shall discuss in the next section, such a non-trivial $\dot b$ 
induces Lorentz- (LV) and CPT-Violating (CPTV)
leptogenesis in models that involve massive right-handed neutrinos (RHN) in their spectra~\cite{decesare,bms1,bms2,bms3,bms4}.

The condensate \eqref{rrt2} also leads to the condensate $\langle b R_{\mu\nu\rho\sigma} \, \widetilde R^{\mu\nu\rho\sigma} \rangle$. The latter remains {\it approximately constant }until the end of inflation, at time $t_{\rm end}$, with  $t_{\rm end} \, H_I \simeq \mathcal N$,
where $\mathcal N$ is the number of e-foldings, which phenomenologically can be taken to be  $\mathcal N \simeq 60-70$~\cite{inflation}, provided 
\begin{align}\label{b0}
\frac{|\overline b(0)|}{M_{\rm Pl}}  \gg \, \sqrt{2\, \epsilon} \,  \mathcal N = \mathcal O(10), \quad \overline b(0) < 0~,
\end{align}
This will lead to a de Sitter contribution to the effective gravitational action~\cite{sola4}
\begin{align}\label{lambda}
\mathcal S_\Lambda  &=
\sqrt{\frac{2}{3}}\,
\frac{\alpha^\prime}{96 \, \kappa} \, \int d^4 x \sqrt{-g} \, \langle \overline b \, R_{\mu\mu\rho\sigma}\, \widetilde R^{\mu\nu\rho\sigma} \rangle  \equiv  -  \int d^4x \, \sqrt{-g} \, \frac{\Lambda (H)}{\kappa^2} \nonumber \\ & \simeq   \int d^4 x \, \sqrt{-g}\, \Big(5.86 \times 10^{-5}  \, \Big(\frac{\mu}{M_s}\Big)^4 \, \sqrt{2\, \epsilon} \,
\Big[\frac{\overline b(0)}{M_{\rm Pl}}\Big] \, H^4 \Big)\,.
\end{align}
where the validity of \eqref{mMs} is understood, and we replaced $H_I$ by $H$, allowing for a mild $t$-dependence of the Hubble parameter, for the sake of comparing with the formulae of the RVM framework.

What we shall do next is precisely to evaluate the equation of state of this fluid, 
by evaluating the total energy ($\rho_{\rm total}$) and pressure ($p_{\rm total}$) densities,
which receive contributions from the KR axion field (superscript $b$), the gravitational anomaly Chern-Simons terms (superscript gCS) and 
the condensate \eqref{lambda} (superscript  ``condensate''): 
\begin{align}\label{stiffvacuumexcit}
\rho_{\rm total} =  \rho^b + \rho^{\rm gCS} + \rho^{\rm condensate} , \quad
p_{\rm total} = p^b + p^{\rm gCS} + p^{\rm condensate},
\end{align}
Using properties of the Cotton tensor, discussed in section \ref{sec:string}, 
and exploiting the ``stiff'' equation of state of the KR axion ``matter'', $\rho_b=p_b$,
we straightforwardly arrive at the following relations~\cite{ms1,ms1b}:
\begin{align}\label{bgCS}
p^b = + \rho^b, \, \, p^{\rm gCS} = \frac{1}{3} \, \rho^{\rm gCS} \,\, \Rightarrow \,\,
p^b + p^{\rm gCS} = \rho^b + \frac{1}{3} \, \rho^{\rm gCS} = -\frac{1}{3}\, \rho^{\rm gCS} = - (\rho^b + \rho^{\rm gCS})>0~,
\end{align}
and 
\begin{align}\label{dScond}
p^{\rm condensate} = - \rho^{\rm condensate} \,< \,0,
\end{align}
which imply a total equation of state for $\rho_{\rm total}$ and $p_{\rm total}$
of RVM (de Sitter-like) type \eqref{rvmeos}. 

It is important to stress that \eqref{bgCS} are valid because of generic properties of the Cotton tensor and the stress tensor of the KR axion, $T_{\mu\nu}^b$, independent of the existence of a GW condensate.  From \eqref{bgCS} we observe that, were it not for the condensate \eqref{lambda}, the gravitational anomalies, due to their negative contributions to the energy density, would make that fluid
behave like "phantom matter", with negative energy density and positive pressure, violating the weak energy condition~\cite{phantmat1,phantmat2}. Nonetheless, the dominance of the condensate term \eqref{lambda} 
in the early Universe, which scales like $H^4$ and is characterised by the standard 
de-Sitter-like equation of state \eqref{dScond}, 
renders the total energy density {\it positive}~\cite{sola4}: 
\begin{align}\label{toten}
0 <  \rho_{\rm total}  \simeq  3\kappa^{-4} \, \Big[ -1.65 \times 10^{-3} \Big(\kappa\, H \Big)^2
+ \frac{\sqrt{2}}{3} \, |\overline b(0)| \, \kappa \, \times {5.86\, \times} \, 10^6 \, \left(\kappa\, H \right)^4 \Big]
\end{align}
under the condition \eqref{b0}. 

The energy density \eqref{toten} also has the form \eqref{rgeh4} of a conventional RVM energy density~\cite{sola,sola2,sola3}, but 
the coefficient 
of the $H^2$ term  is {\it negative}, due to the gravitational anomalous Chern-Simons contributions,
\begin{align}\label{nuinfl}
\nu_{\rm infl} = -1.65 \times 10^{-3}\,.
\end{align}
 Moreover, there is also no evidence for the presence of a non-zero constant $c_0$ in this early RVM-inflationary phase, although such a (positive, cosmological) constant can be generated at late eras of the Universe evolution~\cite{sola4}, as we shall discuss in the next section \ref{sec:postinfl}.

As follows from \eqref{bgCS}, \eqref{dScond}, the fluid obeys the RVM equation of state \eqref{rvmeos}, for $H(t)$ depending (slightly) on cosmic time, in the sense that in such a case, the effective gravitational action, describing the dynamics of the system, contains a de-Sitter (positive-cosmological-constant-like) condensate term \eqref{lambda} plus fluctuations around that, the latter being described by the gCS and KR axion terms in the effective action, which obey a RVM (de-Sitter like) equation of state \eqref{bgCS}. 
Hence, 
although in our analysis above we have assumed initially a constant $H$, in order to compute the
GW-induced anomaly condensate \eqref{rrt2}, 
nonetheless the explicit derivation of an RVM equation of state for this fluid, for non constant $H(t)$, implies 
the emergence of a dynamical RVM inflation, without the need for external inflatons, as a solution to the temporal evolution 
of $H(a)$ within the RVM framework \eqref{evolH}, discussed in section \ref{sec:rvm}.  The only difference is that in our stringy early universe there is no radiation or other matter, only fields from the gravitational multiplet of strings dominate this early era. Hence we set $T_{\mu\nu}^{\rm m}=0$ in \eqref{einst}, and identify $T^{\rm RVM}_{\mu\nu}$ with~\cite{sola4,ms1,ms1b} 
\begin{align}\label{total}
T^{\rm RVM}_{\mu\nu} = T^b_{\mu\nu} +  \sqrt{\frac{2}{3}}\,
\frac{\alpha^\prime}{12 \, \kappa} \, {\mathcal C}_{\mu\nu} + \Lambda (H) \, g_{\mu\nu} 
\end{align}
where $\Lambda (H) $ is the GW condensate \eqref{lambda}. The (total) vacuum stress tensor $T^{\rm RVM}_{\mu\nu}$ is conserved on account of \eqref{bcons}, during the inflationary phase for which $\Lambda \simeq $ constant, since $H \simeq $ constant, but also when one considers the more general RVM case of a time dependent $H(t)$, as a result of the Bianchi identity for the covariant derivative of the Einstein tensor.  

It is important to stress that in our stringy-RVM approach~\cite{ms1,ms1b} we consider the KR axion field contribution to \eqref{total} as a ``vacuum contribution'', in view of the association of this field with the massless gravitational multiplet of the underlying string theory, which in the (phenomenologically relevant) case of superstrings is also the ground state of the string. 
Thus, we have to set the matter/radiation stress tensor $T^{ m}_{\mu\nu} = 0$ in \eqref{einst}, or, equivalently, in the case of ideal FLRW universes we assume here,     
$\rho_m = p_m=0$  for the respective energy and pressure densities. This is perfectly consistent with the RVM evolution, as becomes clear from \eqref{early}, in the sense that there is no appreciable matter/radiation at the early stages of the RVM inflation. In our stringy-RVM approach, chiral matter is assumed to be generated at the end of the inflationary period~\cite{sola4,ms1,ms1b}, as a result of the decay of the running vacuum~\cite{baslima,baslima2}.

Some important comments are now in order regarding the above point. 
In some scenarios for the origin of GW, discussed in the previous subsection \ref{sec:originGW}, we have seen that a pre-RVM inflationary phase is invoked, and even a first (hill-top) inflationary phase (see fig.~\ref{fig:sugra}) can exist. As we have discussed, at the end of the first inflation in dynamically-broken supergravity theories~\cite{houston,houston2}, for instance, which are embeddable in superstring theories, one may have generation of KR axions and gravitino condensates, which are present with some finite densities after the generation of GW from collapsing DW. In the absence of GW, the KR axion has a  `stiff' equation of state $w_b=1$, but the massive gravitino condensate (with mass even close to Planck mass), could be considered as a non relativistic matter with equation of state $w_{\rm gravitino~condensate} =0$. The issue in such scenarios is which of these two kinds of fields dominates the pre-RVM inflationary era. This issue is easily resolved, in the sense that the massive gravitino condensate can easily decay (including among its decay products the massless KR axions), and thus at the end of the pre-inflationary era, i.e. during GW condensation, there is dominance of the massless KR axion (which, as mentioned above, is a field that belongs to the string ground state, and in this sense is considered as part of the RVM ground state rather than relativistic matter).  Such a stiff-era dominance in the very early universe is reminiscent of the stiff-matter scenario of Zeldovich~\cite{stiff}, but there the stiff matter was ordinary baryonic matter,\footnote{Cosmology with abstract stiff matter was also considered in \cite{stiff2}.} in contrast to our stringy case, which is associated with pseudoscalar (KR axion) contributions the string vacuum.  

\begin{figure}[t]
\centering\includegraphics[width=5in]{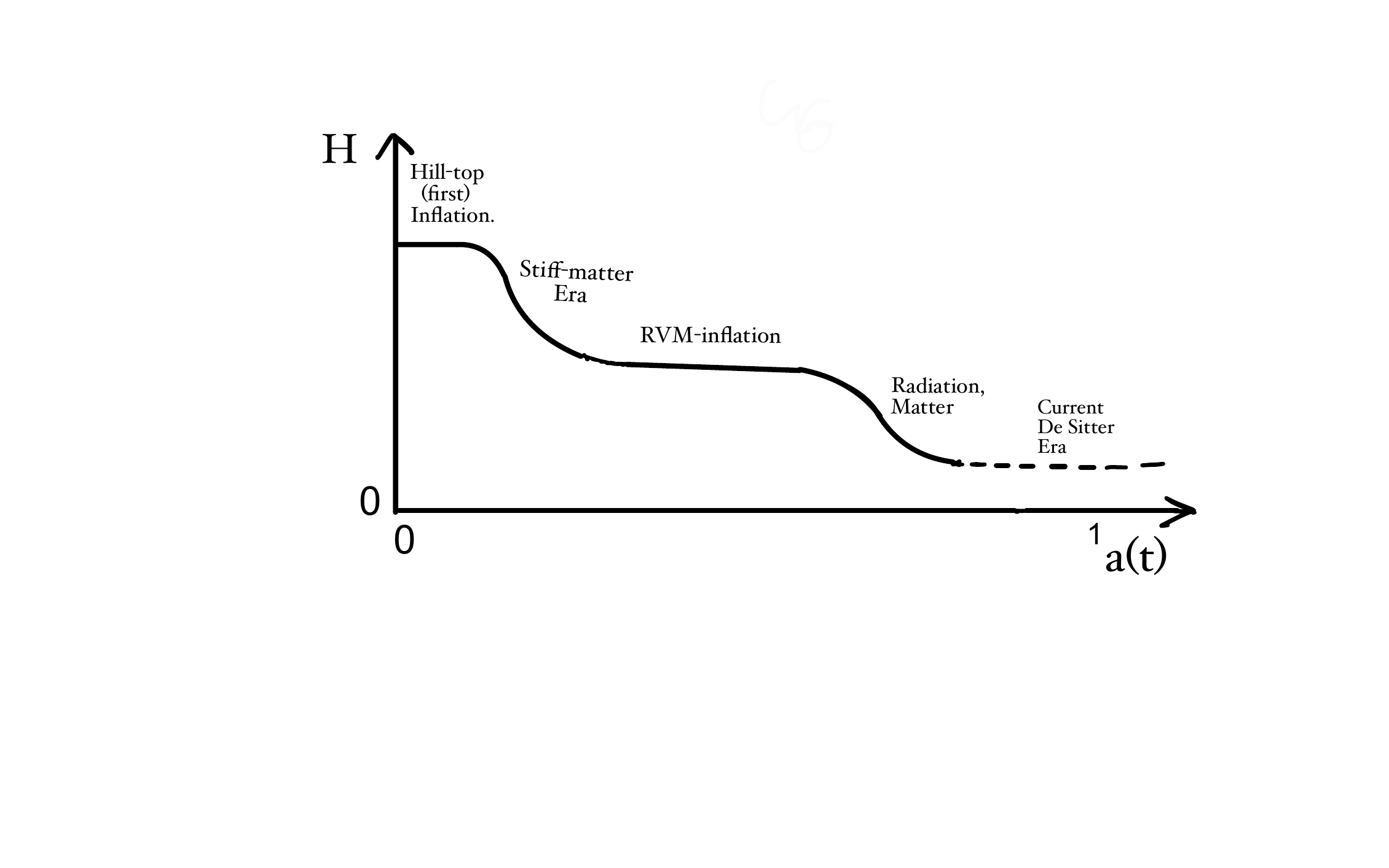}
\vspace{-2cm}
\caption{Diagram of the Hublle parameter ($H$) vs. the scale factor $a(t)$ in the stringy RVM in the scenario with two inflationary epochs, separated by a stiff-KR-axion dominated era. The first (hill-top) inflation, near the Big Bang ($a(t) \to 0$), occurs in models with dynamical supergravity breaking, as a result of gravitino condensation~\cite{hilltop}. The second inflation is of RVM type and is due to GW-induced condensation of gravitational anomalies, which characterise the string-inspired gravitational model at early epochs. In string theory models, due to higher-curvature corrections in the low-energy target-space effective action, an initial singularity of the Universe might be avoided. This is a featured shared with the RVM cosmology. Figure taken from \cite{ms1,ms1b}.}
\label{fig:evolH}
\end{figure}

During the stiff pre-RVM-inflationary era, then, the axionic-type vacuum energy density will scale as 
(the scale factors $a(t)$ below are expressed in units of today's scale factor):
\begin{align}\label{stiffaxion}
\rho_{{\rm massless~KR~axion}} \sim \frac{\rho_{I~{\rm first}}}{a^6},
\end{align} 
where $\rho_I$ determines the value of the density at the exit from the first hill-top inflationary phase. Upon assuming that the onset of the RVM inflation that succeeds the 
stiff era, that is, when GW condensation occurs \eqref{rrt2} and $H$ enters a constant de-Sitter value again $H\simeq H_I$ (see fig.~\ref{fig:evolH}), occurs at cosmic time $t \simeq t_{{\rm RVM~infl~onset}
}$, we may match \eqref{stiffaxion} with 
the RVM density value \eqref{matterearly} : 
\begin{align}\label{condI}
 \frac{\rho_{I~{\rm first}}}{a^6(t_{{\rm RVM~infl~onset}})} 
 \simeq  \frac{3\, H_I^2}{\kappa^2} \, \frac{(1-\nu)}{\alpha} \quad \Rightarrow \quad 
 a(t_{{\rm RVM~infl~onset}}) \simeq \Big(\frac{3\, H_I^2}{\kappa^2} \, \frac{(1-\nu)}{ \rho_{I~{\rm first}}\,\alpha}\Big)^{-1/6} \ll 1\,,
\end{align}
where $H_I$ is the Hubble parameter during the GW-induced RVM second inflation, which can be measured by CMB data \eqref{PlHI}~\cite{Planck}. On account of \eqref{alpha}, 
as well as the fact ({\it cf}. \eqref{toten}, \eqref{nuinfl}) that in our stringy RVM, during the inflatiionary phase, we have $|\nu|=|\nu_{\rm infl}|  \ll 1$, 
the condition \eqref{condI} leads to
\begin{align}\label{rhoinflI}
\rho_{I~{\rm first}} \ll 3 \frac{H_I^2}{M_{\rm Pl}^2} \, M_{\rm Pl}^4  \sim  3 \times 10^{-10} \, M_{\rm Pl}^4,
\end{align}
due to \eqref{PlHI}, which in turn implies subplanckian values for the energy density at the end of the first hill-top inflation, consistently with the transplanckian conjecture.

\subsection{On potential primordial-black-hole effects on GW during RVM Inflation \label{pBHGW}}

Before closing this section we would like to make some speculative but important remarks on potential effects of the primordial black holes on the primordial GW  spectrum in this framework. The RVM-like inflation induced by the anomaly condensation  could in principle change the density and features of the primordial black holes, which in turn could cause an effect on GW. We cannot give a definite answer to this question unless details of the string theory {\it multi-axionic} spectrum~\cite{svrcek}, arising from compactification, are implemented~\cite{ms1b}. Indeed, in such a case, it is possible that some of these extra axions develop periodic instanton-induced potentials during the RVM inflation (par contrast, in our scenario, the KR axion $b(x)$ field could only develop a non-perturbative potential at post inflationary epochs~\cite{ms2}, as we shall discuss in the next section).
Thus, in this multiaxion situation, the effective potential, including the interactions of the string axions $a_i$ $i=1, \dots N$, with the gravitational Chern-Simons terms during the RVM-like inflation, can be represented schematically as 
\begin{align}\label{multiaxion}
V(a_i, b) =  V_0 + \frac{1}{f_b} \, b \, \langle R_{\mu\nu\rho\sigma} \, \widetilde R^{\mu\nu\rho\sigma} \rangle + \sum_i \, \frac{1}{f_{a_i}} \, a_i  \langle R_{\mu\nu\rho\sigma} \, \widetilde R^{\mu\nu\rho\sigma} \rangle + 
\sum_i \Lambda _i(a_i)^4 \, {\rm cos}(\frac{1}{f_{a_i}} a_i ),  
\end{align}
where we assumed the existence of the (constant) gravitational Chern-Simons condensate \eqref{rrt2} during the RVM-like inflation, and the (approximately) constant (for the entire duration of RVM inflation) $V_0 \equiv \langle \overline b (t) \,  R_{\mu\nu\rho\sigma} \, \widetilde R^{\mu\nu\rho\sigma} \rangle $ is the de-Sitter-like condensate ({\it cf.} \eqref{lambda}), with $\overline  b(t)$ the background KR axion \eqref{slowrollbint} (we assume that the other stringy axions do not lead to such contributions).
Here we denoted the background KR axion as $\overline b$ to distinguish it in \eqref{multiaxion} from the quantum KR axion fields $b$, which are viewed as excitations about the background. The quantities $f_b, f_{a_i}$ are the corresponding axion couplings (for the value of $f_b$ see \eqref{sea4}), whilst $\Lambda_i(a_i)$ are appropriate field dependent amplitudes. 

The structures in \eqref{multiaxion} are simplified, given that in realistic situations one may have axion mixing~\cite{ms1b}. However, they are sufficient to demonstrate our main point which concerns the fact that
the presence of the periodic axionic structures in \eqref{multiaxion} might lead~(see, {\it e.g.}, \cite{periodic1,periodic2}) to parametric resonant phenomena for the perturbations of the axions $a_i$, which are amplified significantly at small scales. These can induce fluctuations to the KR axion $b$ field (which does not exhibit oscillatory behaviour), which could be enhanced dramatically~\cite{periodic1}, producing large curvature perturbations. This phenomenon could affect the density of the primordial Black Holes produced during the stringy-RVM inflation. It is possible that one can have an abundant production of such primordial black holes, which, in turn, can affect the primordial GW. In addition, the large fluctuations of the axions $a_i$ could lead~\cite{periodic2} to a dominance of the GW produced during the RVM-like inflation over those produced at the post inflationary radiation era, and the corresponding spectra could have observable effects in interferometers. At present these are speculations, which we hope to pursue further in a future work.

\section{Post-Inflationary era, Kalb-Ramond axions as Dark Matter and Leptogenesis  \label{sec:postinfl}}

At the end of the RVM inflation, the decay of the running vacuum generates radiation and chiral fermionic matter, which itself generates gravitational anomalies. Indeed, chiral fermions in the string effective action couple to the $\mathcal H_{\mu\nu\rho}$-torsion, and by implementing again in the respective path integral the Bianchi identity via the Lagrange multiplier KR axion field $b(x)$, in a similar manner as in the bosonic case, we end up with the following action~\cite{sola4,ms1,ms1b}:
\begin{align}\label{sea6}
S^{\rm eff} &=\; \int d^{4}x\sqrt{-g}\Big[ -\dfrac{1}{2\kappa^{2}}\, R + \frac{1}{2}\, \partial_\mu b \, \partial^\mu b -  \sqrt{\frac{2}{3}}\,
\frac{\alpha^\prime}{96\, \kappa} \, \partial_\mu b(x) \, {\mathcal K}^\mu
\Big] \nonumber \\
&+ S_{\rm Dirac~or~Majorana}^{Free} + \int d^{4}x\sqrt{-g}\, \Big[\Big( {\mathcal F}_\mu + \frac{\alpha^\prime}{2\, \kappa} \, \sqrt{\frac{3}{2}} \, \partial_{\mu}b \Big)\, J^{5\mu}    - \dfrac{3\alpha^{\prime\, 2}}{16 \, \kappa^2}\,J^{5}_{\mu}J^{5\mu}  + \dots \Big] + \dots,
\end{align}
where $S_{Dirac ~or~Majorana}^{Free}$ denote free kinetic terms of (chiral) Dirac or Majorana fermions, which we do not need to specify explicitly, and 
the $\dots$ indicate gauge field kinetic terms, as well as terms of higher order in derivatives, of no relevance to us in this work. In this action, the propagating degrees of freedom are the graviton (and dilaton in general, but here it is considered as constant), the KR axion $b(x)$ (associated with the $\mathcal H$-torsion), and the matter fermions (as well as gauge fields that are generated alongside the chiral matter at the end of the RVM inflation, not exhibited explicitly here). The four-fermion axial-current-current terms are the standard result of integrating out torsion~\cite{shapiro}. The quantity ${\mathcal F}^d  =   \varepsilon^{abcd} \, e_{b\lambda} \,  \partial_a \, e^\lambda_c$, with $e^\lambda_c$ the vielbeins, vanishes for FLRW backgrounds.  In the third term of the second line of \eqref{sea6}, we have performed integration by parts, which lead to the coupling of the KR axion to the divergence of the axial fermion current 
\begin{align}\label{axialcurr}
J^5_\mu = \sum_{j={\rm species}} \overline \psi_{j}\, \gamma^\mu \, \gamma^5 \psi_{j},
\end{align}
where $j$ is a fermion species index. We note that in these early eras, due to high temperatures, the fermions are assumed massless (relativistic matter).
If there are gravitational and chiral anomalies in the theories, this divergence will be linked to them in the standard way (see Appendix)~\cite{anom} :
\begin{eqnarray}
   \label{anom}
J^{5\mu}_{\,\,\,\,\,\,\,\,;\mu}  \!&= \frac{1}{\sqrt{-g}} \, \partial_\mu \Big(\sqrt{-g} \, J^{5\, \mu}\Big) = 
&\! \partial_\mu \Big(\sqrt{-g} \, \frac{\mathcal N}{192 \pi^2} \, \mathcal K^\mu \Big) - 
\frac{e^2\, \mathcal N}{32\pi^2} {\mathbf F}^a_{\mu\nu}\,  \widetilde{\mathbf F}^{a\,\mu\nu}, 
\end{eqnarray}  
where $\mathbf F^a$ the (non-Abelian, in general) gauge field strength, with $a$ gauge-group indices. The quantity $\mathcal N$ indicates the number of chiral stringy degrees of freedom circulating in the chiral-fermion loop, whose precise value depends on the underlying microscopic string model. 

In \cite{sola4,ms1,ms1b} 
we postulated the cancellation of the chiral-fermion-induced gravitational anomalies in \eqref{anom} by the primordial gravitational-anomaly terms that exist in \eqref{sea6},
due to the Green-Schwarz mechanism ({\it cf.} \eqref{GSH}), so that there is no issue with energy conservation of ordinary chiral-fermion matter, after the RVM inflation, and thus standard cosmology is more or less maintained. This is expressed by the fact that the current conservation \eqref{anom2} must imply 
\begin{align}\label{anom2}
& \partial_\mu \Big[\sqrt{-g}\, \Big(  \sqrt{\frac{3}{8}} \frac{\alpha^\prime}{\kappa}\, J^{5\mu}  -  \frac{\alpha^\prime}{\kappa}\, \sqrt{\frac{2}{3}}\,
\frac{1}{96} \, {\mathcal K}^\mu  \Big) \Big]   =   \sqrt{\frac{3}{8}} \, \frac{\alpha^\prime}{\kappa}\, \Big(\frac{\alpha_{\rm EM}}{2\pi}  \, \sqrt{-g}\,  {F}^{\mu\nu}\,  \widetilde{F}_{\mu\nu} + \frac{\alpha_s}{8\pi}\, \sqrt{-g} \, G_{\mu\nu}^a \, \widetilde G^{a\mu\nu} \Big)~,
\end{align}
which, on account of \eqref{sea6} implies in order of magnitude~\cite{ms1,ms1b} $\mathcal N \sim \frac{192\pi^2}{72} \sim 26$.\footnote{Alternatively, to ensure such a cancellation of gravitational anomalies, for arbitrary values of $\mathcal N$, thus not restricting the underlying string model, one may start from a given string theory, with $\mathcal  N$ chiral fermionic degrees of freedom, and modify the Green-Schwarz counterterms 
in the definition of $\mathcal H_{\mu\nu\rho}$, \eqref{GSH}, by a factor $\xi = \frac{72 \, \mathcal N}{192\, \pi^2}$.} 

On the right-hand side of \eqref{anom2} the remaining terms are chiral anomalies, 
of either electromagnetic fields $F_{\mu\nu}$ (with $\alpha_{\rm EM}$ the fine structure ``constant" of electromagnetism), or gluon fields $G_{\mu\nu}^a, a=1, \dots 8$ (with $\alpha_s$ the string-interaction fine structure ``constant" ), 
which do not need to be cancelled, since, as already mentioned, they do not contribute to the stress tensor, and thus they do not affect the energy conservation of ordinary matter in  the epochs after RVM inflation.

\subsection{KR axion mass generation and Dark Matter \label{sec:dmaxion}}

During the QCD era, the second term on the right-hand side of \eqref{anom2} dominates, and is responsible~\cite{ms2}, through non-perturbative (instanton) effects, for the generation of a KR axion pontential
\begin{align}\label{axpot}
V_b = \Lambda_{\rm QCD}^4 \,\Big( 1 -  {\rm cos}(\frac{b}{f_b})\Big), \quad f_b = \frac{1}{\mathcal N} \, \sqrt{\frac{8}{3}}\, \frac{\kappa}{\alpha^\prime} = \frac{1}{\mathcal N} \, \sqrt{\frac{8}{3}}\, \Big( \frac{M_s}{M_{\rm Pl}}\Big)^2 \, M_{\rm Pl},
\end{align}
where $\Lambda_{\rm QCD} \sim \mathcal O(200) $~MeV is the characteristic QCD energy scale,   
and we took into account that \eqref{anom2} implies an axion coupling parameter $f_b = \frac{1}{\mathcal N} \, \sqrt{\frac{8}{3}} \, \frac{M_s^2}{M_{\rm Pl}}$.
This leads to a 
mass for the KR axion field
\begin{align}
m_b \sim \mathcal N \, \sqrt{\frac{3}{8}}\, \Big(\frac{\Lambda_{\rm QCD}}{M_{\rm Pl}}\Big)^2 \, 
 \Big(\frac{M_{\rm Pl}}{M_s}\Big)^2 \,.
\end{align}
Recalling the allowed range of $M_s$, \eqref{msscale}, we obtain a KR axion mass 
$m_b \sim \mathcal N \times 10^{-5}~{\rm eV} \simeq 2.6 \times 10^{-4}$~eV, for $\mathcal N \sim 26$, which lies within the phenomenologically acceptable range for a QCD-type axion~\cite{marsh}.  This massive KR axion 
could also plays the r\^ole of (a component of) DM. In this way, taking into account the association of the KR axion with torsion within the context of the underlying string theory, we obtain a geometric origin of DM in this stringy-RVM framework~\cite{ms1,ms1b,ms3}. In generic models, as we have explained above, the energy scale $\Lambda_{\rm QCD}$ appearing in the non-perturbative KR-axion potential \eqref{axpot} could be an arbitrary scale, to be determined phenomenologically. Such more general situations, where the remaining axions in string theory arising from compactification also appear and might mix with $b$, have been conjectured and discussed briefly in \cite{sola4,ms1,ms1b}. In that case the mass of the KR axion is arbitrary, and can even take on very small values, thus implying ultralight axions, which exhibit a rich phenomenology~\cite{marsh,arvanitaki}.

\subsection{Leptogenesis \label{sec:lepto}}

Another important physical consequence of the stringy-RVM model is that it leads to 
lepton-antilepton asymmetry in the Universe ({\it leptogenesis}) in models which involve sterile right-handed neutrinos in their spectra~\cite{sola4,ms1,ms1b}. The type of induced {\it leptogenesis} is that studied in \cite{decesare,bms1,bms2,bms3,bms4}, which is encountered in the case of a decay of right-handed neutrinos 
in the presence of approximately constant axial-vector backgrounds, coupled to the axial fermionic current, which  {\it spontaneously} violate  Lorentz (and CPT) symmetry.  

Crucial to this effect is the LV and CPTV solution of the KR axion field 
\eqref{slowrollbint}, \eqref{krbeom2}, due to the presence of a GW-induced-anomaly condensate
$\mathcal K^0$. Such an approximately KR-axion background remains undiluted till the end of inflation, as mentioned previously. 
At the exit from inflation in the stringy RVM model the aforementioned assumption of
cancellation of gravitational anomalies during the post inflationary era would imply an equation
of motion for the KR background (stemming from \eqref{sea6}) of the form:
\begin{align}\label{eqmotionpostinfl}
0 &= \frac{1}{\sqrt{-g}}\, \partial_\mu \Big(\sqrt{-g}\,\Big[ \partial^\mu b - \frac{\alpha^\prime}{2\kappa} \sqrt{\frac{3}{2}} \, J^{5\mu} \Big] \Big)=   \frac{1}{\sqrt{-g}}\, \partial_\mu \Big(\sqrt{-g}\, \partial^\mu b \Big) - \frac{\alpha^\prime}{2\kappa} \sqrt{\frac{3}{2}} \, J^{5\mu}_{\,\,\,\,\,;\mu} \nonumber \\ & = 
\frac{1}{\sqrt{-g}}\, \partial_\mu \Big(\sqrt{-g}\, \partial^\mu b \Big) + {\mathcal O}\Big(\frac{\alpha_{\rm EM}}{2\pi} \,  {F}^{\mu\nu}\,  \widetilde{F}_{\mu\nu}\, , \, \frac{\alpha_s}{8\pi}\, \sqrt{-g} \, G_{\mu\nu}^a \, \widetilde G^{a\mu\nu} \Big)~, 
\end{align}
where the last term on the right-hand-side of the last equality, in the second line, of the above equation denotes the chiral anomalies appearing in the anomaly equation of the axial fermion current \eqref{anom2}.  Although the chiral anomalies can be present in the post inflationary epochs of the universe, nonetheless, as we discussed previously, they are dominant at later eras, e.g the QCD epochs, where they are responsible for KR-axion masses, and not during the early radiation era that succeeds the RVM inflation. This implies that immediately after  the RVM inflation the KR-axion classical equation, obeyed by the axion background, reads:
\begin{align}\label{kreomrad} 
\partial_\mu \Big(\sqrt{-g}\, \partial^\mu b \Big) =0. 
\end{align}
For a FLRW universe, with scalar factor during the radiation era scaling with the (cosmic) temperature as $a(T) \sim T^{-1}$, Eq.~\eqref{kreomrad} has the solution for the $b(t)$ KR-axion field
during the early radiation era 
\begin{align}\label{krbsol}
\dot b_{\rm early~radfiation} \sim a^{-3}(T) \sim T^{3}.
\end{align}
In the context of the stringy RVM,  one can match the solution \eqref{krbsol} at a temperature $T$ with the value \eqref{slowrollbint} at the exit from the RVM inflation, where the temperature can be taken~\cite{sola4} to be of order of the 
Gibbons-Hawking temperature~\cite{GH} of the (approximately) de Sitter space time,  $T_{\rm GH} = H_I/(2\pi)$. Then,  one obtains:
\begin{align}\label{bscale2}
\dot{b}_{\rm early~radfiation} \simeq  3.5 \times 10^{11} \, M_{\rm Pl}^2 \, \Big(\frac{T}{M_{\rm Pl}}\Big)^3,
\end{align}
which can be used for the lepton-asymmetry computations.

During the short-period of leptogenesis, compared to the universe evolution, and for the sufficiently high temperatures that the lepton-asymmetry generation takes place ({\it e.g.} freezeout temperatures $T = T_D \simeq m_N \ge 10^5 $~TeV
in the model of \cite{decesare,bms1,bms2,bms3,bms4} ({\it cf.} \eqref{tdmn}, below), with $m_N$ the mass scale of the decaying sterile neutrino, which constitutes a typical temperature range for such models), one may view the background \eqref{bscale2} as {\it approximately constant}. In such a case, the sterile neutrino part of the action \eqref{sea6} reads (in the case of a single species of a Majorana sterile massive neutrino, $N$, which suffices to generate leptogenesis, and fits our purposes here):
\begin{align}\label{smelag}
\mathcal{L}= {\mathcal L}_{\rm SM} + i\overline{N}\, \gamma^\mu\, \partial_\mu \, N-\frac{m_N}{2}(\overline{N^{c}}N+\overline{N}N^{c})-\overline{N}\gamma^\mu\, B_\mu \, \gamma^{5}N-\sum_f \, y_{f}\overline{L}_{f}\tilde{\phi}^dN+ {\rm h.c.}
\end{align}
where h.c.  denotes hermitian conjugate, ${\mathcal L}_{\rm SM}$ denotes the Standard Model (SM) Lagrangian,
$\tilde \phi$ is the SU(2)-``dual'' of the Higgs field  $\phi$ ($\tilde{\phi}^d_i \equiv \varepsilon_{ij}\phi_j~, \, i,j=1,2,$ SU(2) indices),
 and $L_{f}$ is a lepton (doublet) field of the SM sector, with $f$ a generation index, $f=e, \mu, \tau$, in a standard notation for the three SM generations; $y_f$ is a Yukawa coupling, which is non-zero and provides a non-trivial (``Higgs portal'') interaction between the RHN and the SM sector, used in the seesaw mechanisms~\cite{seesaw,seesaw2,seesaw3,seesaw4,seesaw5} for generation of SM neutrino masses. The  quantity $B_\mu$   is an axial background, given by
\begin{align}\label{background}
B_\mu = M_{\rm Pl}^{-1} \, \dot{\overline b}\, \delta_{\mu0}\,,
\end{align}
on account of \eqref{sea6}. The background is assumed to be approximately constant due to our previous discussion. Under this assumption, the Lagrangian \eqref{smelag} acquires the form of a Standard-Model-Extension (SME) Lagrangian~\cite{sme,smebounds}, with the background axial vector violating (spontaneously) Lorentz and CPT symmetries.

In the context of the Lagrangian \eqref{smelag}, with \eqref{background}, \eqref{bscale2}, one can calculate the decay rates of the Majorana massive neutrino into standard model massless leptons and Higgs particles, including charged Higgs excitations, given that we consider temperatures much higher than the electroweak phase transition~\cite{decesare,bms1,bms2,bms3,bms4}. The lepton asymmetry 
$\Delta L$ is generated due to the different decay rates between such decays, and those where the products are the corresponding antiparticles: 
\begin{align}\label{channels}
&{\rm Channel} ~I: \,  N \rightarrow \ell^{-}h^{+}~, ~ \nu \, h^{0},  \nonumber \\
&{\rm Channel ~II}: \, N \rightarrow \ell^{+}h^{-}~,~  \overline \nu \, h^{0},
\end{align}
where $\ell^\pm$ denote charged leptons, $h^\pm$ charged Higgs excitations, $h^0$ is the neutral Higgs (or Higgs particle of the SM after electroweak symmetry breaking) and $\nu\, (\overline \nu$) are the light neutrinos (antineutrinos) of the SM. For small $|M_{\rm Pl}^2 \dot b | \ll 1$, which  characterises our stringy RVM,  the lepton asymmetry is calculated, to leading order in $\dot b$, to be~\cite{bms2}:
 \begin{align}\label{lepto}
 \frac{ \Delta L^{TOT}(T=T_D)}{s} \sim  q\,  \dfrac{B_{0}(T_D)\, m_N^2}{T_D^3} \sim \, q\, 3.5 \times 10^{11} \,
 \Big(\dfrac{m_N}{m_{\rm Pl}}\Big)^2, \quad q > 0,
 \end{align}
where $s$ is the entropy density of the universe, $T_D$ is the freezeout temperature, assumed to be of order of the sterile neutrino mass, $T_D \sim m_N$, in the model of \cite{decesare}, which we adopt here, and $q$ is a numerical factor of order $\mathcal O(10)$, which is due to theoretical uncertainties in the semi-analytical method (Pad\'e approximants) used in the calculation of $\Delta L$~\cite{bms2}.

This lepton asymmetry can be communicated to the baryon sector via sphaleron processes in the SM sector of the  model~\cite{baryogen,baryogen2,baryogen3}, which violate Baryon(B) and lepton(L) numbers,   
B+L, but conserve their difference B-L. In this way, requiring that the lepton asymmetry \eqref{lepto} reproduces the observed baryon asymmetry in the Universe~\cite{Planck}: 
\begin{align}\label{bau}
\frac{n_b - n_{\overline b}}{n_b + n_{\overline b}} \simeq (8.4-8.9) \times 10^{-11} , 
\end{align}
for temperatures $T > 1$~GeV (with $n_{b(\overline b)}$ the density of baryons (antibaryons) in the Universe), we obtain 
\begin{align}\label{tdmn}
T_D \simeq m_N \sim 10^7~{\rm GeV}.
\end{align}
This magnitude of $m_N$ is compatible~\cite{sola4} 
with the seesaw mechanisms~\cite{seesaw,seesaw2,seesaw3,seesaw4,seesaw5} and Higgs-mass stability~\cite{Hstab,Hstab2}. Once again, given that this leptogenesis is exclusively due to the LV and CPTV KR-axion background \eqref{slowrollbint}, which is linked to the antisymmetric-tensor-field torsion in the underlying string theory, one obtains a {\it geometric interpretation} of the {\it matter-antimatter} asymmetry in the cosmos, should this scenario be realised in nature~\cite{ms3}. 

\section{Modern Era: RVM-like deviations from $\Lambda$CDM and alleviation of cosmological-data tensions  \label{sec:modern}}

The modern era of the stringy-RVM is the less understood from a formal point of view. We still do not understand what plays the r\^ole of the current-era cosmological constant, as the data indicate~\cite{Planck}. As discussed in \cite{sola4}, in the modern epoch, the chiral electromagnetic U(1) anomalies that are generated after the exit from the RVM inflation, during the late radiation era, may survive until today, in the sense that the terms $F_{\mu\nu} \widetilde F^{\mu\nu}$ in \eqref{anom2} are non zero. Such cosmic electromagnetic field terms, leading to a KR-axion electrodynamics may contribute to the vacuum energy density terms of the form $\nu_0 H_0^2$, where $H_0(t)$ the Hubble parameter in the current era, where the coefficient $\nu_0 > 0$, as in the conventional RVM. However, there might be other contributions to the vacuum energy density, which we are not aware of at present. Terms of $H_0^4$ are in any case not dominant in the current epoch, so such terms do not constitute the focus of our attention. As already mentioned, we also still do not understand the microscopic origin of a current-era cosmological constant , if any. In the context of string theory/brane theory there might be several scenarios which could lead to a de-Sitter-like (positive) cosmological constant in the current four-dimensional Universe, see {\it e.g.} \cite{rizos,rizos2}, but we will not discuss them in this article. 

\subsection{Highlights of the RVM phenomenology in the current era \label{sec:rvmphenocurr}}

What we shall discuss, though, is how the RVM framework in the modern era, can contribute to observable, in principle, deviations from the $\Lambda$CDM~\cite{sola5,sola6,tsiapi1,tsiapi2,sola11}, and most importantly alleviation of the current cosmological-data tensions $H_0$ and $\sigma_8$~\cite{sola7,sola8,sola9}.  In fact, as we shall discuss now, a slightly modified version of the conventional RVM, has the capacity of alleviating {\it simultaneously} the $H_0$ and $\sigma_8$ tensions~\cite{sola10}.

Let us first remind the reader that in the modern era, the dominant terms in the RVM energy density of the cosmological vacuum are up and including the $H^2$ terms of \eqref{rgeh4}:
\begin{align}\label{rLRVM0}
\rho_{0\,\rm RVM} \simeq  \frac{3}{\kappa^2} \, \left(c_0 + \nu_0 H_0^{2} \right), \qquad c_0\,, \, \nu_0 > 0~,
\end{align}
where, in a standard notation in cosmology, a subscript ``$0$'' indicates present-day quantities.
First we remark that fitting \eqref{rLRVM0} to the plethora of the cosmological data~\cite{Planck} 
leads to the conclusion that ({\it cf.} also \eqref{nucurr})~\cite{sola5,sola6}:
\begin{align}\label{nuc0}
\nu_0 = \mathcal O(10^{-3}) > 0~, \qquad   3\kappa^2 \, c_0 = \mathcal O(10^{-122})  > 0~.
\end{align}
As already discussed in section \ref{sec:rvm}, the presence of the $H_0^2$ term in \eqref{rLRVM0}
leads to observable in principle deviations from $\Lambda$CDM, in the sense that there is a 
different scaling \eqref{modern} of the Hubble parameter today compared to the prediction of the $\Lambda$CDM paradigm.

However there is an even more important r\^ole of the RVM, which allows, as already mentioned, for a {\it simultaneous alleviation} of the $H_0$ and $\sigma_8$ tension~\cite{sola10}. 
To this end, one needs to consider a variant of RVM, called type-II RVM, which allows for a cosmic-time dependence of the gravitational constant 
\begin{align}\label{timdk}
\kappa^2 \to \kappa^2(t) = \kappa^2 \, \varphi^{-1}(t)~,
\end{align} 
where $\varphi (t)$ is a phenomenological function of the cosmic FLRW time $t$. This implies
the modification:
 \begin{align}\label{rLRVM0timdk}
\rho_{0\,\rm RVM~type-II} \simeq  \frac{3}{\kappa^2} \, \varphi(t) \, \left(c_0 + \nu_0 H_0^{2} \right), \quad c_0, \, \nu_0 > 0~,
\end{align}
which should be compared agains the plethora of the available cosmological data~\cite{Planck,sola10}.
It should be stressed that the function $\varphi(t)$ is not the Brans-Dicke (BD) field, and the type-II RVM is defined only through the above modification \eqref{rLRVM0timdk}, without ascribing a dynamical r\^ole for the function $\varphi(t)$. 
In fact a comparison on the phenomenology of both the BD cosmology and the type-II RVM, insofar as the alleviations of both $H_0,\sigma_8$ has been made in \cite{sola10}, with the conclusion  that the type-II RVM variant is superior in the simultaneous alleviation of both tensions,
provided one chooses a function $\varphi (t)$ with a mild (l{\it ogarithmic}) dependence on the cosmic time. 

\subsection{Modified stringy RVM due to Quantum-Gravity Corrections? \label{sec:qgcorr}}

It is interesting to see whether such modifications are allowed within our stringy-RVM framework.
This question is still pending a rigorous proof, but below we can make some remarks on existing theoretical evidence~\cite{ms1,ms3,ms4} towards it. The evidence is provided in the work of \cite{houston,houston2} on dynamically-broken $N=1$ supergravity~\cite{N1sugra} at one-loop in a de Sitter background, {\it i.e.} in a space-time with cosmological constant $\Lambda > 0$. The formalism for the one-loop quantisation of the theory, which is an example of a weak (perturbative) QG treatment, follows the pioneering work of \cite{fradkin}. This example, as we have discussed in section \ref{sec:gw}, is quite relevant for our stringy-RVM scenario, first because such models can be embedded in superstring theory, but more specifically because it may characterise the early stages of our string-inspired cosmology, providing a natural origin of GW.

We should stress~\cite{ms4} at this point, for completion, that the background space-time used in 
\cite{fradkin,houston,houston2}
is a local de Sitter space time, with metric described by the following invariant element
\begin{align}\label{SDS}
 ds^2= \left(1-2\frac{M}{\bar{r}} - \frac{\Lambda}{3}\, \bar{r}^2 \right) c^2 dT^2 - \left(1-2\frac{M}{\bar{r}} - \frac{\Lambda}{3}\, \bar{r}^2 \right)^{-1} d\bar{r}^2
          -\bar{r}^2 (d\bar{\theta}^2 +  \rm{sin}^2\bar{\theta} d\bar{\phi}^2)
\end{align}
in de Sitter-Schwarzschild coordinates. In the case $M=0$, as required by the isotropy and homogeneity of space, which is the one used as a background for the $N=1$ supergravity example, there exists a coordinate transformation~\cite{lanczos}
\begin{align}\label{lanczostr}
x^\mu &\equiv \{ c\,T, {\bar r}, {\bar \theta}, {\bar \phi} \} \, \rightarrow  \, x^{\prime \, \mu} \equiv \{c\,t, r, {\bar \theta}, {\bar \phi} \}  \, \,\,{\rm (comoving~frame)} \, : \nonumber \\
t &= T + \frac{1}{2 \, H} \, {\rm ln} \Big(1 - H^2 \, {\bar r}^2 \Big), \quad
r = \frac{{\bar r}}{\sqrt{1 - H^2 \, {\bar r}^2 }}\, e^{-H\, T} = {\bar r} \, e^{-H t},  \quad H^2 \equiv \frac{\Lambda}{3} \, > \, 0~.
\end{align}
where $t$ and $r$ denote co-moving-frame time and radial space coordinates, respectively, which 
maps the metric \eqref{SDS} to a standard cosmological de Sitter space-time:
\begin{align}\label{inflation}
ds^2 = c^2 \, dt^2 - a(t)^2 \big[ dr^2 + r^2 d\Omega^2 \big], \quad d\Omega^2 = d\bar{\theta}^2 +  {\rm sin}^2\bar{\theta} \, d\bar{\phi}^2~,
\end{align}
where $a(t)=e^{\sqrt{\frac{\Lambda}{3}}\, t} \equiv e^{H\, t} $ ($H=$ constant), is the exponentially expanding scale factor of the de-Sitter/FLRW (inflationary) Universe. The above result is valid for every non-negative value of the cosmological constant $\Lambda \ge 0$.  The effective action is invariant under the general coordinate transformation \eqref{lanczostr}, and this makes the connection of the approach of \cite{fradkin,houston,houston2} to our cosmological model.

The one-loop effective action of dynamically-broken $N=1$ supergravity  in a Euclidean (E) path-integral formalism, in a local de Sitter background, with one-loop-renormalised (positive) cosmological constant $\Lambda > 0$, is given by~\cite{fradkin,houston,houston2}:
\begin{align}\label{effactionl2}
    \Gamma^{(\rm E)} \simeq S_{\rm cl}-\frac{24\pi^2}{\Lambda^2 }\big(\alpha^F_0+\alpha_0^B
    + \left(\alpha^F_{1}+ \alpha^B_{1}\right)\Lambda 
    +\left(\alpha^F_{2}+ \alpha^B_{2}\right)\Lambda^2+\dots\big)~,
\end{align} 
in a specific gauge~\footnote{We do  not discuss here issues of gauge invariance. This is a complicated issue, and since the present case  constitutes a weak QG effective action, we refer the reader to gauge invariant formulations of QG, like, for instance, the Batalin-Vilkovisky formalism~\cite{BV}.},
where the superscripts $B (F)$ refer to terms arising from integration of massless (quantum) gravitons (Bosonic (B) degrees of freedom (d.o.f.)) (and gravitinos (Fermionic (F) d.o.f.). The quantity $\sigma_c < f$,~\cite{houston2} denotes the value of the gravitino condensate field at the minimum of its one-loop effective  potential (see fig.~\ref{fig:sugra}) 
and 
$S_{\rm cl}$ denotes the classical supergravity action with tree-level (bare) cosmological constant $\Lambda_0 < 0$ 
\begin{align}\label{l0tree}
	 \frac{\Lambda_0}{\kappa^2} 	
	 \equiv \sigma_c^2  -f^2 ~,
\end{align}
where $f$ is the energy scale of dynamical breaking of supergravity (and also global supersymmetry). We stress that the bare cosmological constant $\Lambda_0$ 
is necessarily \emph{negative}, given that (unbroken) supergravity (local supersymmetry), which characterises the action $S_{\rm cl}$, is incompatible with de Sitter vacua~\cite{houston,houston2,N1sugra}. The one-loop renormalised cosmological constant 
$\Lambda$, one the other hand, is {\it positive}, due to quantum corrections (see discussion below, Eq.~\eqref{LL1}), and this is compatible with the case of dynamically-broken supergravity~\cite{wittendbs}.

On taking into account that the Euclidean de Sitter volume is~\cite{fradkin}  
\begin{align}\label{desittervol}
24\pi^2/\Lambda^2 \rightarrow \int d^4x \sqrt{\widehat g_{\rm E}}~,
\end{align} 
we observe that the effective action \eqref{effactionl2} can be written in a covariant form  (with the notation that hatted quantities denote those evaluated in the de-Sitter background space-time):
\begin{align}\label{effactionl3}
\Gamma^{\rm (E)} \simeq&-\frac{1}{2\kappa^2} \int d^4 x \sqrt{\widehat g_{\rm E}}
\left[\left(\widehat R-2\Lambda_1 \right)  +\alpha_1 \, \widehat R+
\alpha_2 \, \widehat R^2\right]~,
\end{align}
where we used the fact that the curvature
scalar in the (Euclidean) four-dimensional de Sitter space time is given by: 
\begin{align}\label{rl}
\widehat R = 4\Lambda,
\end{align}
or 
\begin{align}\label{cosdS}
\widehat R = 12 \overline H^2, \quad \overline H = {\rm constant}
\end{align}
in the cosmological de Sitter case, with a constant Hubble parameter $\overline H$, of interest to us here.

The remaining quantities in \eqref{effactionl3} are given by~\cite{houston2} 
    \begin{align} \label{LL1}
\Lambda_{1}=-\, \kappa^2 \,
\left(-\frac{\Lambda_0}{\kappa^2}+\alpha_0^F+\alpha_0^B\right)~,
    \end{align}
which expresses one-loop corrections to the bare cosmological constant, with    
\begin{align}\label{a0}
		\alpha_0^{F} &= {\kappa}^4 \, \sigma_c^4 \, \Big(0.100\,  \ln \left( \frac{{\kappa}^2 \, \sigma_c^2}{3 \mu_\tau^2}\right) + 0.126 \Big)~, \nonumber \\
		 				\alpha_0^{B} &= \kappa ^4 \, \left(f^2-\sigma_c^2\right)^2 \left(0.027 - 0.018 \ln \left(\frac{3 \kappa ^2 \left(f^2-\sigma_c^2\right)}{2 \mu_\tau^2}\right)\right)~,
		\end{align}
		and
\begin{align}\label{alphasugra}
         \alpha_1=\frac{\kappa^2}{2}\left(\alpha^F_1+\alpha^B_1\right)~,\quad
        \alpha_2=\frac{\kappa^2}{8}\left(\alpha^F_2+\alpha^B_2\right)~,
    \end{align}
    where
\begin{eqnarray}\label{aif}
\alpha^F_1&=& 0.067\, \kappa^2 \sigma_c ^2  -0.021\,
\tilde\kappa^2 \sigma_c ^2 \, {\rm ln}
\left(\frac{\Lambda}{\mu_\tau^2}\right)
 +  0.073\, \kappa^2 \sigma_c ^2 \, {\rm ln}
\left(\frac{\kappa^2\sigma_c^2}{\mu_\tau^2} \right)~,
\nonumber \\
\alpha^F_{2}&=& 0.029 + 0.014\, {\rm ln} \left(\frac{\kappa^2\sigma_c^2}{\mu^2}\right) 
-0.029\, {\rm ln} \left(\frac{\Lambda}{\mu_\tau^2}\right)~, \nonumber \\
\alpha^B_1&=& -0.083 \Lambda_0 + 0.018\, \Lambda_0 \, {\rm ln} \left(\frac{\Lambda }{3 
\mu_\tau^2}\right)  + 
0.049\, \Lambda_0\,  {\rm ln} \left(-\frac{3 \Lambda_0}{\mu_\tau^2}\right)~, \nonumber \\ \alpha^B_{2} &=& 0.020 +  0.021\, {\rm ln} \left(\frac{\Lambda }{3 \mu_\tau^2}\right) - 
0.014\, {\rm ln} \left(-\frac{6 \Lambda_0}{\mu_\tau^2}\right)~.
\end{eqnarray}
The replacement of $\Lambda$ in all the above expressions by the scalar curvature, \eqref{rl}, is understood. The quantity $\mu_\tau^2$ (with dimensions of mass-squared) is a UV ultraviolet cutoff on the proper-time $\tau$, which regularises UV divergences~\cite{houston,houston2,fradkin}, that is, small $\mu_\tau^2$ values correspond to the UV regime of the theory, whilst large $\mu_\tau^2$ values correspond to the infrared (IR). 
In the work of \cite{houston}, supergravity breaks dynamically at a large scale $\mu_\tau^2$ close to Planck scale~\cite{houston}, which allows us to set  from now on
\begin{align}\label{muPl}
\mu_\tau^2 \sim M^2_{\rm Pl} = \kappa^{-2}.
\end{align}
In the dynamically-broken-supergravity phase, the gravitino and its condensate acquire large masses, which can be well above the grand-unification scale, even close to Planck scale for our purposes~\cite{ms1,ms2,ms4,rvmsugra}, since we want supergravity to be broken well before the RVM inflation (see fig.~\ref{fig:evolH}). This  can be arrange for appropriate values of the scale $f$.

The conjectural (at this stage) interpretation that the ${\rm ln}\Lambda $ terms in the effective action of the dynamically-broken supergravity theory, are  not simple coefficients of curvature terms, dependent on the renormalised $\Lambda$, but can themselves be viewed as covariant 
curvature scalars not only in de Sitter background \eqref{rl}, but also away from the de-Sitter values, and therefore can be considered as non-polynomial contributions to the effective action \eqref{effactionl3}, implies that the Quantum supergravity effective action now has terms 
$\widehat R^{n}\, {\rm ln}(\kappa^2 \widehat R)$, $n=1,2$, on account of \eqref{muPl}.\footnote{The reader should notice that our modified gravity has a Minkowski flat limit, as the curvature scalar $R \to 0$, and in this respect it has to be contrasted with the purely ${\rm ln}R$ gravity suggested in \cite{nojiri}, whose terms grow with small curvatures. In this respect, our proposal, as well as that of \cite{houston,rvmsugra} shares features with the modified gravity with $R^2$ terms of \cite{tanmoy4}. In this latter reference, antisymmetric tensor 
fields are also considered. In our (3+1)-dimensional setting, in our case these would correspond to our KR axions, which are present in the modern eras, as discussed above.}  

Moreover, in the broken supergravity phase, the gravitino and its condensate, being superheavy excitations in our stringy-RVM case~\cite{ms1,ms2,ms4,rvmsugra}, examined here, are integrated out in the path-integral, given that they will lead to terms in the effective action suppressed by the Planck scale, leaving only terms of the massless degrees of freedom, which eventually will constitute the effective action \eqref{sea4}. 
We observe from \eqref{alphasugra},\eqref{aif} that such logarithmic $\widehat R^{n}\, {\rm ln}(\kappa^2\,\widehat R)$ is also the result of integrating out massless graviton fluctuations. Thus, apart from the relevance of the $N=1$ supergravity example to the early stages of our  stringy-RVM universe~(see fig.\ref{fig:evolH}), as described above in section \ref{sec:gw}, we may also encounter such modified relativity (bosonic) effective actions in the modern era. That is, based on the supergravity example, we may conjecture~\cite{ms1,ms3,ms4} that the result of integrating out massless graviton fluctuations in 
the effective action, which  describes the gravitational dynamics of the post inflationary stringy RVM Universe, until the current era, leads to weak QG corrections, which are given by adding to the standard Einstein-Hilbert Lagrangian term one-loop corrections of the form (we analytically continue back to a Minkowski-signature space-time from now on):
\begin{equation}\label{1looplagr}
\delta \mathcal L^{\rm 1-loop}_{\rm quant.~grav.} =  - \sqrt{-\widehat g}\, \Big[ \widetilde \alpha_0 + \widehat R \Big(c_1 + c_2\, {\rm ln}(\frac{1}{12}\kappa^2 \widehat R)\Big) \Big] + \dots 
\end{equation}
where $\widetilde \alpha_0$ plays the r\^ole of a one-loop induced cosmological constant. 
>From the supergravity example~\cite{houston,houston2} we have seen that $\widetilde \alpha_0 >0$, and 
that the constant coefficients $c_i$ assume  either the form $c_i \propto \kappa^2 \mathcal E_0, \, {\rm or} \, c_i \propto \kappa^2 \mathcal E_0 \, {\rm ln}(\kappa^4 |\mathcal E_0|)$, $ i=1,2,$ ({\it cf.} \eqref{aif}) with $\mathcal E_0$ a bare (constant) vacuum energy density scale.  
The ellipses $\dots $ in \eqref{1looplagr} denote terms of quadratic and higher order in $\widehat R = 12 H^2$ ({\it cf.} \eqref{cosdS}), which are subdominant in the current epoch $(H = H_0)$), when the universe enters again a de Sitter phase. The structures \eqref{1looplagr} appear generic for weak QG corrections about de Sitter backgrounds~\cite{fradkin}, as appropriate for the current era of the universe. We may therefore conjecture that the corections \eqref{1looplagr} can lead to a modified version of the stringy RVM discussed so far, thus playing a r\^ole in the current-era phenomenology.\footnote{The presence of $H^4 {\rm ln}(\kappa^2 H^2)$ QG-induced terms in the early Universe ({\it cf.} $\widehat R^2 {\rm ln}(\kappa^2 \widehat R)$ terms in \eqref{effactionl3}), which are dominant during RVM inflation, are not affecting our considerations in this work. Indeed, such terms are subdominant 
compared to the GW-induced $H^4$ terms in the condensate \eqref{toten}, for 
$\kappa^4 |\mathcal E_0| < 1$, as required by the transplanckian conjecrture, which the bare scale $\mathcal E_0$ is assumed to satisfy. Thus our conclusions on RVM inflation remain unaffected.}

Indeed, considering the graviton equations stemming from the one-loop corrected effective Lagrangian, we easily observe~\cite{ms3,ms4} that the correction terms \eqref{1looplagr} 
imply corrections to the effective stress-energy tensor in the current era of the form~\cite{ms3,ms4}, 
\begin{equation}\label{1loopenden}
\delta \rho_0^{\rm vac}  =  \frac{1}{2}\widetilde \alpha_0 + 3 (c_1-c_2) \, H_0^2 + 3 c_2 \, H_0^2 \, {\rm ln}(\kappa^2 H_0^2) + \dots~, 
\end{equation}
where the $\dots$ denote subleading terms proportional to $(\dot H_0)^2, \ddot H_0$, which are negligible in the current epoch, during which the universe enters once again a de-Sitter phase. 
 
The total stress energy tensor is obtained by adding \eqref{1loopenden} to \eqref{rLRVM0}, assuming the existence of a bare cosmological constant. 
It is important to notice that the supergravity prototype~\cite{houston,houston2,rvmsugra} indicates that the one-loop correction (dark-energy-type) term $\frac{1}{2} \widetilde \alpha_0$ is constant, {\it independent} of ${\rm ln} H^2$ terms. This will lead to some crucial differences from the standard type II RVM \eqref{rLRVM0timdk}, used in \cite{sola10} for the simultaneous alleviation of  the $H_0$ and $\sigma_8$ tensions, which is characterised only by a mild cosmic-time $t$ dependence of an effective gravitational constant, and as such it necessarily contains a time-dependent dark energy term.  
The phenomenology of the QG-modified stringy RVM, 
as far as the current-epoch tensions in the cosmological data are concerned, will be examined elsewhere.

\subsection{Brief comparison of the stringy-RVM with other theories in contorted geometries \label{sec:contorted} }

We conclude this section, with a brief comparison of the main cosmological applications of the stringy RVM, described above, with those of some other cosmological models  in contorted geometries that exist in the contemporary literature.

We commence our discussion with the $f(\overline R)$ cosmology theories with torsion~\cite{tors1}, with 
$\overline R$ the generalised curvature scalar which includes the torsion. In that work, the torsion is considered as purely geometric, without being associated with a spin fluid. 
In fact, it is demonstrated that only for the case $f(\overline R)=\overline R^2$, the torsion has a non-trivial effect on the vacuum, leading to an accelerated expansion. It is the trace of an appropriately modified torsion that plays a crucial r\^ole in inducing an inflationary phase for the Universe, in the sense of the pertinent cosmological equations in the model leading to a scale factor that depend exponentially on that trace. Specifically, in the notation 
of \cite{tors1}, if $S_{\mu\nu}^{\,\,\,\,\,\,\,\lambda}$ denotes the torsion, then the modified torsion used in that work is defined as $T_{\mu\nu}^{\,\,\,\,\,\,\,\lambda} = S_{\mu\nu}^{\,\,\,\,\,\,\,\lambda} + \delta_\mu^{\,\,\lambda} \, S_\nu - \delta_\nu^{\,\,\lambda} \, S_\mu$. The temporal component $T_0$ of its trace $T_\mu \equiv T_{\mu\nu}^{\,\,\,\,\,\,\,\nu} =  -2 S_\mu, \, \mu=0, \dots 3,$ is linked to an exponential expansion of the Universe, with scale factor 
$a(t) = a_0\,\exp\Big(-\frac{T_0}{3} \, t \Big)$ or $a(t) =a_0\,  \exp\Big(-\frac{T_0}{3} \, t + A_0 \, \exp(\frac{T_0}{3}\, t)\Big)$ depending on which case of the solution is satisfied, with $a_0,A_0$ arbitrary constants (expansion of the universe occurs in both cases for $T_0 < 0$).

Par contrast, in our stringy-RVM scenario, it is the GW-induced condensate of gravitational anomalies which leads to a de Sitter era, of RVM type, in a gravity theory with Einstein-Hilbert terms $R$ in the Lagrangian. Moreover, the torsion in our case is totally antisymmetric, hence its trace vanishes, but such a torsion is associated with a pseudoscalar  field, the KR axion, which couples to the anomaly terms, and in this sense, the r\^ole of the torsion in inducing inflation is crucial. This last feature is therefore shared by the model of \cite{tors1}, where it is the torsion that is equivalent to a scalar degree of freedom that leads to inflation. However, as we have discussed above, the stringy-RVM has an additional feature which distinguishes it from the framework of \cite{tors1}, that of the potential r\^ole of the KR axion (torsion) as DM component, under some conditions of mass generation due to QCD instantons in the presence  of chiral anomalies of the colour gauge group SU(3)$_c$ in the post-inflationary era.

The cosmology with spin of ref.~\cite{tors2} associates torsion with a fluid with spin (chiral fermions), in contrast to the case of \cite{tors1}. In contrast also to our case, the cosmology of \cite{tors2}, makes use of the Einstein-Cartan-Holst action, {\it i.e.} including the term \eqref{holst}, 
whilst our cosmology made use of generalised  topological invariants \eqref{nieh3}, which are linear combinations of the Nieh-Yan invariant and the anomaly terms, due to the Green-Schwarz mechanism. An important feature of the approach of \cite{tors2} is the exploitation of the four-fermion interactions which arise as a generic feature of theories with torsion, 
upon solving the pertinent equations of motion, thus expressing torsion classically in terms of the chiral fermions. In this way, theories with such non-vanishing fermionic torsion are equivalent to adding these fermion self-interactions to a torsion-free gravity. By choosing appropriately the chiral-spinorial matter in the model, 
these fermion self-interactions can come up with arbitrary coefficients, of arbitrary sign, that is the self interaction can be attractive or repulsive. 
The upshot is that several cosmological models can emerge with different characteristics, among which cosmologies with a bounce, in which an initial singularity is absent. In contrast, in our approach, the coefficient and signature of the four-fermion interactions \eqref{sea6} are fixed, the former depending on a combination of the string and Planck scales, and the latter being such that the four-fermion interactions are {\it necessarily} repulsive. We note in passing that the existence of attractive four fermion interactions may lead to torsion fermion condensates, which can contribute to the dark energy sector, but also break dynamically CPT symmetry~\cite{popl1,popl2}, thus providing a source for matter-antimatter asymmetry in the universe, although such issues have not been discussed in \cite{tors2}.

In this latter respect, in our stringy-RVM model, as we have seen,  matter-antimatter asymmetry can also be linked to a CPTV and LV torsion condensate, in the sense of leptogenesis taking place in such backgrounds, which however in our case is due exclusively to gravitational anomalies, due to a primordial-GW condensation.
The absence of initial singularity in our string-inspired  model is also associated, 
at the level the pertinent low-energy effective gravitational action,
with the contributions of (an infinity of) higher curvature terms., hence the early universe physics is different. Moreover, the totally antisymmetric torsion, that characterises the string model, is equivalent at a quantum level to a fully fledged axion field, which arises as a Lagrange multiplier implementing the Bianchi identity constraint \eqref{modbianchi2}. This last feature has not been discussed in the context of the model of \cite{tors2}, however, we believe that by requiring torsion conservation to all orders in perturbation theory in such models, one should be able to see the association  of an axion with the totally antisymmetric part of the fermionic torsion of the model of \cite{tors2} (which has more than a totally antisymmetric component). This happens in the model of QED with torsion~\cite{kaloper}, which we discuss in the Appendix. 

Finally, we close the discussion of this review by mentioning that there is another approach to torsion, that of teleparallel gravity (and cosmology)~\cite{tele,heis1,heis2,heis3,tele2}, in which torsion mimics the r\^ole of the gravitational field. Several aspects of the ordinary gravity are shared in this teleparallel approach, for instance one can get cosmological inflation, of rather long duration in $f(T)$ teleparallel theories, with $T$ the so-called torsion scalar, or late-time acceleration in the Universe,  whilst signals that mimic the standard gravitational waves characterise such theories. Moreover, the appearance of non-singular bounce cosmologies is another feature  of these models. As also mentioned in the introduction of the review, models including the so-called non-metricity scalar $Q$ (defined as an appropriate contracted combination of the non-metricity tensor $Q_{\alpha\mu\nu} \equiv \overline D_\alpha g_{\mu\nu}$, with $\overline D_\alpha$ the gravitational covariant derivative with torsion, which, in contrast to the torsion and the Riemann curvature tensors,  depends on {\it both} the metric and the connection), {\it e.g.} $f(Q)$ gravities~\cite{heis4,heis5}, are currently being investigated in the literature, insofar as their cosmological and black-hole-physics aspects are concerned.

Our stringy model is different in many respects from such theories, given that gravity is necessarily included in the spectrum of strings, corresponding to the spin-2 
massless excitation of the bosonic string gravitational multiplet, whilst torsion is associated with the 
spin-1 antisymmetric tensor field, as we have discussed, and in this sense is associated with an extra field added to the usual curvature-based formulation of gravity. Par contrast, in teleparallel theories of gravity, torsion is inherently linked  to the massless spin-2 degrees of freedom of the theory (graviton), and replaces the curvature formulation.
Moreover, in our approach, gravitational anomalies play a crucial r\^ole in inducing inflation. It would be interesting to investigate further the r\^ole of gravitational anomalies in teleparallel theories of gravity, for instance, along the lines of~\cite{andrade,andrade2}. There is an interesting question to be answered in this respect, namely whether teleparallel theories of gravity can give rise to RVM cosmologies, like in the case of our stringy-RVM reviewed above. This remains open at present. \\{\it Affaire \`a suivre....}

%%%%%%%%%%%%%%%%%%%%%%%%%%%%%%%%%%%%%%%%%%
\acknowledgments

The author wishes to thank the Guest Editors of the Special Issue: {\it Beyond Riemannian Geometry in
Classical and Quantum Gravity},  Dr. Marco Danilo Claudio Torri, Dr. Christian Pfeifer and Dr. Nicoleta Voicu
for their kind invitation to contribute to this special issue, and for organising such a high-level and stimulating volume. This work is supported in part by 
the UK Science and Technology Facilities  research Council (STFC) under the research grant ST/T000759/1. 
N.E.M.  also acknowledges participation in the COST Association Action CA18108 ``{\it Quantum Gravity Phenomenology in the Multimessenger Approach (QG-MM)}''.

\appendix

\appendix

\numberwithin{equation}{section}

\setcounter{equation}{0}

\section{Torsion formalism in an instructive example: Quantum Electrodynamics in a contorted geometry \label{sec:app}}

Before formulating the Quantum Electrodynamics (QED) in the presence of torsion, we shall first review basic elements of the torsion formalism. 
Let us consider a (3+1)-dimensional contorted space time in the sense of Einstein-Cartan. In differential-form language~\cite{eguchi}, which we shall use below for notational convenience, the torsion two-form is defined as:\footnote{A differential $p$-form in (3+1)-dimensional spacetime is defined as 
$\mathbf{\mathcal A^{(p)}} \equiv \mathcal A_{\mu_1 \dots \mu_p} \, \mathbf dx^{\mu_1} \wedge \dots \wedge \mathbf dx^{\mu_p}$, where $\mu_p = 0, \dots 3$ are world indices, and $\wedge$ denotes exterior (antisymmetric) product. Thus, the tensor $\mathcal A_{\mu_1 \dots \mu_p} \equiv n! \,\mathcal A_{[\mu_1 \dots \mu_p]} $, which defines the components of the $p$-form, is totally antisymmetric in its lower indices ($[\dots ]$ denotes antisymmetrisation in the respective indices, with normalization $\frac{1}{n!}$).}
\begin{align}\label{deftors}
\mathbf T^a = \mathbf d \mathbf e^a + \overline \omega^a_{\,\,\,b} \wedge \mathbf e^b \equiv \overline {\mathbf D}\, \mathbf e^a \ne 0,
\end{align}
where $\mathbf d= \partial_\mu \, dx^\mu$ is the exterior derivative one-form,  $\mathbf e^a = e^a_{\,\,\,\mu} (x) \,\mathbf d x^\mu$ is the vielbein one-form, 
and  $\overline \omega^a_{\,\,\,b}$ the generalised spin connection with torsion, 
 which in the presence of a non-zero torsion \eqref{deftors} is independent from the vielbeins (Palatini formalism of general relativity). Above, Greek indices refer to the spacetime manifold, with metric $g_{\mu\nu}(x)$ (of signature $(+,-,-,-)$ in our conventions), and Latin  indices refer to the tangent-space at a point $x$ of spacetime, with flat Minkowski metric $\eta^{ab} =(1, -1,-1,-1) $. We have by definition
 \begin{align}
 g_{\mu\nu} = \eta_{ab}\, e^a_{\,\,\,\mu} \, e^b_{\,\,\,\nu},  \qquad g^{\mu\nu} = \eta^{ab}\, E^\mu_{\,\,\,a} \, E^\nu_{\,\,\,b},   \quad \mu, \nu=0, \dots 3, \, \, a, b = 0, \dots 3,
 \end{align}
with $E^\mu_{\,\,\,a}$ the inverse vielbeins, $E^a_{\mu} \, e^\mu_{\,\,\, b}= \delta^a_{\,\,\,b}$ and 
$E^a_{\mu} \, e^\nu_{\,\,\, a}= \delta_\mu^{\,\,\,\nu}$.

The symbol 
\begin{align}\label{covdertor}
\overline{\mathbf D}^a_{\,b}  \equiv \delta^a_{\, b} \mathbf d + \overline{\omega}^a_{\,b} \wedge 
\end{align}
denotes the gravitational covariant derivative operator in the presence of torsion. The generalised spin-connection (with torsion) can be split into the standard Riemannian part $\omega_{\mu\,\,b}^a$ without torsion, and the {\it contorsion} $K_{\mu\,\,b}^a  \equiv  K_{c\,\,b}^a \, e^c_{\, \mu}$, with $\mathbf K_{ab}=-\mathbf K_{ba}$:
\begin{align}\label{contor}
\overline \omega_{\mu\,\,b}^a = \omega_{\mu\,\,b}^a + K_{\mu\,\,b}^a .
\end{align} 
Hence, we may write 
\begin{align}\label{covdertorsplit}
\overline{\mathbf D}^a_{\,b}  =  \mathbf D^a_{\, b} + K^a_{\,b} \wedge  
\end{align}
where $\mathbf D^a_{\, b} \equiv  \delta^a_{\, b} \mathbf d + \omega^a_{\,b} \wedge $ is the torsion-free gravitational covariant derivative, defined in terms of the 
the Riemannian connection, which satisfies the metricity postulate
\begin{align}
\mathbf D e^a \equiv \mathbf d e^a + \omega^a_{\,b} \wedge e^b =0.
\end{align}
It is this relation in Riemannian spaces that 
eventually allows the torsion-free Christofell symbol ($\Gamma^\mu_{\nu\rho} = \Gamma^\mu_{\rho\nu}$) to be expressed in terms of derivatives of the metric and the metric itself.

>From \eqref{deftors} and \eqref{contor} we easily obtain in differential form notation: 
\begin{align}\label{tcont}
\mathbf T^a = \mathbf K^a_{\,\,b} \, \mathbf e^b \, \Rightarrow \, T^a_{\,\,bc} = - 2 K^a_{\,\,[bc]}
\equiv  K^a_{\,\,bc} - K^a_{\,\,cb}, \, \quad K_{abc} = \frac{1}{2}\Big(T_{cab} - T_{abc} - T_{cba} \Big),
\end{align}
with the symbol $[ \dots ]$ denoting antisymmetrisation of the respective indices, as defined above.

The torsion two-form \eqref{deftors} in coordinate basis is related to the torsion tensor $T^\lambda_{\,\,\,\mu\nu} $ via $T_{\mu\nu}^a = e^a_\lambda \, T^\lambda_{\,\,\,\mu\nu} $, where 
\begin{align}
T_{\,\,\,\mu\nu}^\lambda = \overline{\Gamma}^\lambda_{\,\,\,\mu\nu} - \overline{\Gamma}^\lambda_{\,\,\,\nu\mu} \ne 0,
\end{align}
and $\overline{\Gamma}^\lambda_{\,\,\,\mu\nu}$ is the generalised spin connection, not symmetric in its lower indices in the presence of torsion, which appears in the generalised  Riemann curvature tensor (for our conventions on the metric see \eqref{riemann} in the main text)
\begin{align}\label{riemann2}
\overline R^\lambda_{\,\,\rho\mu\nu} = E^\lambda_{\,\,a}\, e^b_{\,\,\rho} \, \overline R^a_{\,\,b\mu\nu} = \partial_\mu \overline \Gamma_{\,\,\rho\nu}^\lambda + \overline \Gamma^\lambda_{\,\,\sigma\mu}\, \overline \Gamma^\sigma_{\,\,\rho\nu} - (\mu \leftrightarrow \nu),
\quad \lambda,\mu,\nu,\rho = 0, \dots 3~,
\end{align} 
with the (generalised) curvature two form given by  
\begin{align}\label{curv2form}
\overline{\mathbf R}^a_{\,\, b} = \mathbf d \overline \omega^a_{\,\, b} + \overline \omega^a_{\,\, c} \wedge \overline \omega^c_{\,\, b} = \frac{1}{2} \overline R^a_{\,\, b c d} \mathbf e^c \, \mathbf e^d = \frac{1}{2} \overline R^a_{\,\, b \mu\nu} \mathbf d x^\mu \wedge \mathbf d x^\nu, 
\quad \overline \omega^a_{\,\, b} \equiv  \overline \omega^a_{\mu\,\, b} \, dx^\mu.
\end{align}
This can be expressed in terms of the Riemannian  curvature $\mathbf R^a_{\,\,b} $ and contorsion $\mathbf K^a_{\,\,b}$ two forms as:
\begin{align}\label{curv2formriem}
\overline{\mathbf R}^a_{\,\, b} &= \mathbf R^a_{\,\,b} + \mathbf D \, \mathbf K^a_{\,\,b} + 
\mathbf K^a_{\,\,c} \wedge \mathbf K^c_{\,\,b}~,
\end{align}
with $\mathbf R^a_{\,\,b} =  \mathbf d \omega^a_{\,\, b} + \omega^a_{\,\, c} \wedge \omega^c_{\,\, b}$, the Riemannian curvature two form without torsion. 

We observe from \eqref{curv2formriem}, that the effective action based on the generalised scalar curvature has the form (in differential form language),  
\begin{align}\label{gract}
\mathcal S_{\rm G} = \frac{1}{2\kappa^2} \int d^4 x \overline{\mathbf R}_{ab} \wedge \star 
(\mathbf e^a \wedge \mathbf e^b ) = \frac{1}{2\kappa^2} \int d^4 x \Big( \mathbf R_{ab}  + 
\mathbf K_{ac} \wedge \mathbf K^c_{\,\,b} ) \wedge \star (\mathbf e^a \wedge \mathbf e^b),
\end{align}
where the $\star$ denotes the Hodge dual~\cite{eguchi}, defined through its action on a $p$ form in (3+1)-dimensional spacetime as: 
\begin{align}\label{hodge}
\star (\mathbf e^{a_1} \dots \mathbf e^{a_p} ) = 
\frac{1}{(4-p)!} \epsilon^{a_1 \dots a_{p}}_{\qquad  \,\, c_1 \dots c_{4-p}} \, \mathbf e^{c_1} \wedge \dots \wedge \mathbf e^{c_{4-p}}~,
\end{align}
with $\epsilon^{a_1 \dots a_{p}}_{\qquad  \,\, c_1 \dots c_{4-p}} $ the Levi-Civita fully antisymmetric symbol in the (Minkowski-flat) tangent space. 
In \eqref{gract} we took into account that the covariant derivative middle term on the right-hand side of \eqref{curv2formriem} contributes a vanishing boundary term, which vanishes on account of our assumption that fields and their derivatives vanish at the boundary of the space-time manifold.
Thus we arrive at the generic result that the contorted gravitational action \eqref{gract} is quadratic in the contorsion, and does not contain derivatives of the contorsion tensor. This is a generic result in Einstein-Cartan contorted geometries, which allows the contorsion to be integrated out exactly in the respective path integral. 

Let us explore the consequences of this by considering now the Quantum Electrodynamics (QED) 
of a charged Dirac fermion field, $\psi (x)$, for concreteness, in a spacetime with torsion. The pertinent action involving the spinors reads (we follow the conventions of \cite{kaloper}):
\begin{align}\label{qedaction}
\mathcal S_{\rm QED}^{\rm torsion} &= \mathcal S_{\rm Maxwell} + \mathcal S_{\psi}^{\rm torsion} \, ,
\nonumber \\
\mathcal S_{\rm Maxwell} & = -\frac{1}{4} \, \int d^4 x \, \sqrt{-g}\, F_{\mu\nu}(A) \, F^{\mu\nu}(A), \quad F_{\mu\nu}(A)  = \partial_\mu A_\nu - \partial_\nu A_\mu, \nonumber \\ 
\mathcal S_{\psi}^{\rm torsion} &= \int d^4 x \, \sqrt{-g}\,  \Big(\overline \psi \, \frac{i}{2} \,\gamma^\mu  \overline{\mathcal D}_\mu \, \psi - \frac{i}{2} (\overline{\mathcal D}_\mu \overline \psi) \, \gamma^\mu \, \psi - m\,\overline \psi \, \psi \Big)\,
\end{align}
where $\overline{\mathcal D}_\mu = \overline D_\mu - i\, e\, A_\mu(x)$, with $e$ the electric charge (taken here to be that of the electron), is the gauge/gravitational convariant derivative in a space-time with torsion, 
$A_\mu(x)$ is the photon field, and, following \cite{kaloper}, we defined the Maxwell tensor, $F_{\mu\nu}$ without torsion (so that the Christoffel connection parts of the Riemannian gravitational covariant derivative acting on $A_\mu$ cancel in the definition of the Maxwell tensor due to symmetry reasons). The gravitational covariant derivatives $\overline D_\mu$ acting on spinors are defined as:
\begin{align}\label{spinor}
\overline{\mathbf D} \psi = \mathbf d \psi - \frac{i}{2} \, \overline \omega_{ab} \, \sigma^{ab} \, \psi\, ,  \qquad  \sigma^{ab} = \frac{i}{2} \Big[ \gamma^a\, , \,  \gamma^b\, \Big],
\end{align} 
and $\gamma^a$ are the Dirac $\gamma$ matrices in the (flat Minkowski) tangent space (where the indices are raised and lowered with the Minkowski metric $\eta^{ab}$). A similar expression characterises the Riemannian-spacetime covariant derivative, $D_\mu$, but with the replacement of the generalised spin connection one-form $\overline \omega_{ab}$ by the Riemannian one without torsion, $\omega_{ab}$, 
\begin{align}\label{spinor2}
\mathbf D \psi = \mathbf d \psi - \frac{i}{2} \, \omega_{ab} \, \sigma^{ab} \, \psi\,.
\end{align}
On making use of the property of the product of three $\gamma^a$ matrices in tangent space
$$  \gamma^a \, \gamma^b \, \gamma^c = \eta^{ab}\, \gamma^c + \eta^{bc}\, \gamma^\alpha - \eta^{ac}\, \gamma^b - i \epsilon^{abcd} \, \gamma^5 \, \gamma_d, $$
and taking into account \eqref{contor}, we may write the spinor action $\mathcal S_\psi$ in \eqref{qedaction} in the form:
 \begin{align}\label{spinoract2}
\mathcal S_{\psi}^{\rm torsion} &= \int d^4 x \, \sqrt{-g}\,  \Big(\overline \psi \, \frac{i}{2} \,\gamma^\mu  D_\mu \, \psi - \frac{i}{2} (D_\mu \overline \psi) \, \gamma^\mu \, \psi - m\,\overline \psi \, \psi + e \, A_\mu(x) \, \overline \psi \, \gamma^\mu \, \psi   - \frac{1}{4} \, \epsilon^{abcd} \,  \overline \psi \, \gamma^5 \, \gamma_d \, \psi \, K_{abc} \Big).
\end{align}
We see from \eqref{spinoract2} and \eqref{tcont}, then, that it is only the fully antisymmetric part of the (con)torsion, 
\begin{align}\label{antisym}
T_{[abc]}= - 2 K_{[abc]}, 
\end{align}
that couples to fermionic matter, and, in fact, to the axial fermion current \eqref{axialcurr} in the contorted QED action. If we define
the torsion three form 
\begin{align}
\mathbf T = \frac{1}{3!} T_{abc}\, \mathbf e^a \wedge \, \mathbf e^b \wedge \mathbf e^c, 
\end{align}
and its dual one form
\begin{align}\label{sform}
\mathbf S = \star \mathbf T, \quad {\rm with~components} \quad S_d = \frac{1}{3!} \, \epsilon^{abc}_{\quad \,\, d} \, T_{abc}\,, 
\end{align}
then we may write the contorsion term of \eqref{spinoract2} (last on the right-hand side) as~\cite{kaloper}
\begin{align}\label{torsionsform} 
\mathcal S_\psi \ni -\frac{3}{4}\,  \int d^4x \, \sqrt{-g}\, S_\mu \, \overline \psi \, \gamma^\mu \, \gamma^5  \, \psi = 
-\frac{3}{4}\, \int \mathbf S \wedge \star \mathbf J^5
\end{align}
with the axial current one form $\mathbf J^5 = J^5_\mu \, \mathbf dx^\mu $, with components
given by \eqref{axialcurr} (for a single fermion species $\psi$, in this case).

We may decompose the contorsion into its totally antisymmetric part and the rest, denoted by $\widehat K_{abc}$~\cite{kaloper},
\begin{align}\label{decomptorsion}
K_{abc} = \frac{1}{2} \epsilon_{abcd}\, S^d + \widehat K_{abcd}\,,
\end{align}
which, on account of 
\eqref{antisym}, \eqref{sform}, implies that the gravitational action with torsion \eqref{gract} can be written as 
\begin{align}\label{gravactS}
\mathcal S_{\rm G} = \frac{1}{2\kappa^2} \int d^4 x \sqrt{-g} \, \frac{1}{2\, \kappa^2} \,  \Big(R  + \widehat \Delta \Big) + \frac{3}{4\, \kappa^2} \, \int \mathbf S \wedge \star \mathbf S, 
\end{align}
with $\widehat \Delta = \widehat K^\lambda_{\,\,\,\,\mu\nu}\, \widehat K^{\nu\mu}_{\,\,\,\,\,\,\lambda} 
- \widehat K_{\,\,\,\,\,\,\,\nu}^{\mu\nu}\, \widehat K_{\mu\lambda}^{\,\,\,\,\,\,\lambda} $.

When we combine the QED action \eqref{qedaction} with the gravitational one \eqref{gravactS}, we observe that the torsion components are non-propagating fields, as no derivatives of them appear,
and hence they can be integrated exactly in the path integral. The path-integration of the $\mathbf S$ form, dual to the totally antisymmetric part of the torsion, which is the only one coupled to matter, yields four fermion {\it repulsive} integrations~\cite{kaloper}, which are characteristic of the 
Einstein-Cartan theory:
\begin{align}\label{4fermi}
-\frac{3\, \kappa^2}{16} \, \int \mathbf J^5 \wedge \star \mathbf J^5\,.
\end{align}
However, the QED with torsion is still not complete because the axial current $\mathbf J^5$ is {\it anomalous}, that is, it is not conserved at one loop, although {\it classically} it is conserved, as follows from the Euler-lagrange equations of motion stemming from the combined actions \eqref{gravactS}, \eqref{spinoract2} (taking into account \eqref{torsionsform}). 

The $\mathbf S$ equations of motion imply
\begin{align}\label{SJ}
\mathbf S = \frac{1}{2}\, \kappa^2 \, \mathbf J^5.
\end{align}
On the other hand, the fermion equations of motion:
\begin{align}\label{fermioneq}
i \, \gamma^\mu \mathcal D_\mu \, \psi  - m \, \psi = \frac{3}{4} S_\mu \, \gamma^\mu\, \gamma^5 \, \psi,
\end{align}
with $\mathcal D= D_\mu - i \, e\, A_\mu (x)$ the gauge/gravitational covariant derivative, in the absence of torsion, imply that $\mathbf J^5$ is covariantly 
conserved {\it classically} in the {\it massless} limit 
\begin{align}\label{massless}
D_\mu J^{5\mu} \to 0 \quad {\rm for} \quad m \to 0~,
\end{align}
implying, due to \eqref{SJ}:
\begin{align}\label{classconsaxial}
\mathbf d \star \mathbf S = 0 \quad {\rm for} \quad m=0\,, 
\end{align}
which in components leads to\footnote{Note that, in view of \eqref{sform},  the ordinary exterior derivative $\mathbf d$ can be replaced by the Riemannian gravitational covariant derivative $D_\mu$ without torsion, for symmetry reasons.}
\begin{align}\label{compaxial}
D_\lambda \, S^\lambda = 0.
\end{align}

In the presence of a finite fermion mass, on the other hand, one has {\it classically}:
\begin{align}\label{compconsaxialcurr} 
D_\mu J^{5\,\mu } -  2\, i\, m \,\overline \psi \, \gamma^5 \, \psi = 0.
\end{align} 
In what follows we shall concentrate on the case of {\it massless} charged chiral fermions, since in this work we are interested in the physics of the early Universe, where the chiral fermions are produced relativistically after inflation (due to high temepratures), and hence are considered as massless (see discussion around Eq.~\eqref{sea6}, in section \ref{sec:postinfl}). 

In the quantum theory, the conservation law \eqref{massless} breaks down due to chiral anomalies, both in the gauge and gravitational sectors (see also \eqref{anom} in the main text, section \ref{sec:postinfl})~\cite{anom,kaloper}: 
\begin{align}\label{anomalyeq}
D_\mu J^{5\mu} = \frac{e^2}{8\pi^2} F^{\mu\nu}  \, \widetilde F_{\mu\nu} \, - \frac{1}{192\, \pi^2} \overline R^{\mu\nu\rho\sigma} \, \widetilde{\overline R}_{\mu\nu\rho\sigma} \equiv \mathcal G(\overline \omega, A)\,,
\end{align}
where the $\widetilde{(\dots)}$ denotes dual tensors, defined as $\widetilde F_{\mu\nu} \equiv  \frac{1}{2} \sqrt{-g} \, \epsilon_{\mu\nu\rho\sigma} \, F^{\rho\sigma} $, 
 $\widetilde R_{\mu\nu\rho\sigma} \equiv  \frac{1}{2} \sqrt{-g} \, \epsilon_{\mu\nu\alpha\beta} \, R^{\alpha\beta}_{\quad \rho\sigma}$, where $\varepsilon_{\mu\nu\rho\sigma} = \sqrt{-g} \,\epsilon_{\mu\nu\rho\sigma}$ is the gravitationally covariant  Levi-Civita tensor density, 
with $\epsilon_{\mu\nu\alpha\beta}$ the Minkowski flat Levi-Civita symbol. The result \eqref{anomalyeq} is the one-loop chiral anomaly (for one chiral fermion species), but this is an exact result in field theory~\cite{anom}.

At this point, we note that it can be shown~\cite{Hull,mavromatos} (see also \cite{mavsark}) that any torsion dependence on the generalised curvature in the gravitational part of the chiral anomaly \eqref{anomalyeq} can be removed  by adding appropriate renormalization-group counterterms of Green-Schwarz type. This implies that the right-hand-side of \eqref{anomalyeq} depends actually only on the Riemannian part of the connection, and hence the Riemannian curvature tensors, 
\begin{align}\label{notoranom}
\mathcal G(\overline \omega, A) \quad \stackrel{({\rm via~addition~of~counterterms})}{\mathbf \Rightarrow} \quad  \mathcal G(\omega, A)\,.
\end{align}

We next remark that the presence  of the one-loop chiral anomaly \eqref{anomalyeq} implies that as we go to higher loops in our massless QED in contorted geometries, the dual of the totally antisymmetric componnt of the torsion will no longer be conserved, so the right-hand-side of \eqref{classconsaxial} (or, equivalently \eqref{compaxial}) is no longer zero, but is proportional to the chiral anomaly $\mathcal G(\overline \omega, A)$ \eqref{anomalyeq}. To avoid this inconsistency, we may add apporpriate counterterms, order by order in  perturbation theory, to remove such anomalous contributions, so as to ensure \eqref{classconsaxial} at each order. This would imply conservation of the torsion charge in the {\it quantum theory}:
\begin{align}\label{concharge}
Q_{\rm torsion}= \int \star \, \mathbf S.
\end{align}
In a path-integral formalism, then, the above procedure implies that we may add the delta-functional constraint $\delta \Big(\mathbf d \star \mathbf S \Big)$ in the respective path integral, and express the $\delta$-functional via a Lagrange multiplier pseudoscalar (axion-like) field $\varphi(x)$ (we use 
form language for notational brevity)~\cite{kaloper}:
\begin{align}\label{pathintegralqed}
\mathcal Z &\propto \int \mathcal D S \, \delta\Big(\mathbf d \star \mathbf S\Big) \, \exp\Big(i \int \Big[\frac{3}{4\, \kappa^2} \, \mathbf S \wedge \star \mathbf S - \frac{3}{4} \mathbf S \wedge \star \mathbf J^5 \Big]\Big) \nonumber \\ & = \int \mathcal D S \, \mathcal D \varphi \, \exp\Big(i \int \Big[\frac{3}{4\, \kappa^2} \, \mathbf S \wedge \star \mathbf S - \frac{3}{4} \mathbf S \wedge \star \mathbf J^5  + \varphi \, \mathbf d \star \mathbf S \Big]\Big),
\end{align} 
where $\mathcal D (\dots)$ denotes the path-integral measure, and we suppressed the other path integrations over the metric and the other torsion components, as they are not relevant for our arguments in this Appendix. We observe from \eqref{pathintegralqed} that the (non-propagating) 
torsion $\mathbf S$ can be integrated exactly, leading, after appropriate integrations by parts, to
\begin{align}\label{finalaxionpi}
\mathcal Z &\propto \int D b \, \exp\Big(- i \int \Big[ \frac{1}{2} \, \mathbf d \, b  \wedge \star \mathbf d b  + \frac{1}{f_b} \mathbf d b  \wedge \star \mathbf J^5  + \frac{1}{2\, f_b^2} \mathbf J^5 \wedge \star \mathbf J^5 \Big]\Big) \nonumber \\ &= \int D b \, \exp\Big(- i \int \Big[ \frac{1}{2} \, \mathbf d \, b  \wedge \star \mathbf d b  - \frac{1}{f_b}  b\, \mathcal G(\omega, A)  + \frac{1}{2\, f_b^2} \mathbf J^5 \wedge \star \mathbf J^5 \Big]\Big), \qquad f_b \equiv  \Big(\frac{3\kappa^2}{8}\Big)^{-1/2}\,,
\end{align} 
where  
$b = \Big(\frac{3}{2\, \kappa^2}\Big)^{-1/2} \, \varphi $ is a canonically normalised pseudoscalar field, and in the second term of the exponent of the right-hand side of \eqref{finalaxionpi}, we performed  integration by parts, and made use of the anomaly equation \eqref{anomalyeq} and the fact \eqref{notoranom}.

Thus, we see that the presence of torsion in the {\it quantum theory} of contorted QED is equivalent to the introduction of a fully dynamical (propagating) {\it massless} pseudoscalar (axion-like) field $b(x)$, without potential. The reader should notice the natural appearance of the (standard) interaction of the axion field with the chiral anomalies, but with a specific axion coupling $f_b$, depending on the gravitational constant.\footnote{As we have discussed in the main text, section \ref{sec:string}, 
a similar situation characterises the (totally antisymmetric) $\mathcal H$-torsion
 string theory, where the corresponding Lagrange multiplier field, implementing the Bianchi identity \eqref{modbianchi2} in the corresponding path integral \eqref{pathintegral}, is the KR string-model-independent axion~\cite{string,string2,svrcek}, also denoted there by $b(x)$. In the string case the action coupling depends on appopriate combinations of the gravitational constant $\kappa$ and the string Reggie slope $\alpha^\prime$ \eqref{pathintegral}, \eqref{sea6}.} This is the correct treatment of a quantum torsion in this case, which as we see from \eqref{finalaxionpi} also leads to the characteristic {\it repulsive} axial-current-current four-fermion inetractions (see also \eqref{4fermi}) of an Einstein-Cartan theory). We stress once more that, as becomes evident from the above analysis, the coefficient and sign of these four-fermi terms is {\it fixed} for a given theory. 

It is important at this sttage to compare briefly the above results with other analyses of 
torsional models, like the ones of 
\cite{mercuri}, \cite{mercuri2}, \cite{taveras2}, \cite{taveras3}, with which the above analysis is in broad agreement, as we shall explain below.  We have made a similar comparison for the string-inspired torsional model in the main text, in section \ref{sec:string}.
To this end, we note that the last term in the exponent 
on the right-hand side of \eqref{pathintegralqed}, $\int \varphi \, \mathbf d \star \mathbf S $ is, on account of \eqref{sform}, nothing other that the Nieh-Yan topological invariant \eqref{nieh}, with the Lagrange multiplier field $\varphi (x)$ playing the r\^ole of a space-time dependent Barbero-Immirzi (BI) parameter, promoted to a dynamical field (the canonically-normalised, shift-symmetry respecting, axion field $b$, which the totally antisymmetric component of the torsion corresponds to).

On the other hand, in \cite{castel,taveras}, which were the first works to promote the BI parameter  to a dynamical field, the starting point is the so-called Holst action \eqref{holst}, which by itself is {\it not} a topological invariant, in contrast to the Nieh-Yan term \eqref{nieh}. The work of \cite{castel,taveras} deals with matter free cases. If $\gamma (x)$ represents the BI field, the Holst term now reads (in form language)
\begin{align}\label{holstform} 
S_{\rm Holst} = \frac{1}{2\, \kappa^2} \,  \int \overline \gamma(x) \, \mathbf e^a \wedge \mathbf e^b \wedge \overline{\mathbf R}_{ab},
\end{align}
where $\overline{\mathbf R}_{ab}$ is the curvature two-form, in the presence of torsion, and we used the notation of \cite{taveras} for the inverse of the BI field $\overline \gamma(x) = \gamma^{-1}(x)$, to distinguish this case from the KR axion $b(x)$ in our string-inspired one. 
The analysis of \cite{castel,taveras} showed that the gravitational sector results in the action
\begin{align}\label{gravactholst}
\mathcal S^{\rm eff}_{\rm grav+ Holst+BI-field} = 
&=\; \int d^{4}x\sqrt{-g}\Big[ -\dfrac{1}{2\kappa^{2}}\, R + \frac{3}{4 \kappa^2 \, ( \overline \gamma^2 + 1) }\, \partial_\mu \overline \gamma \, \partial^\mu \overline \gamma \Big]
\end{align}

Coupling the theory to fermionic matter~\cite{freidel,rovelli,holstferm} can be achieved by introducing a rather generic non-minimal coupling parameter $\alpha$, for massless Dirac fermions in the form
\begin{align}\label{fermion}
\mathcal S_F = \frac{i}{12} \int \epsilon_{abcd} \mathbf e^a \wedge \mathbf e^b \wedge \mathbf e^c \wedge \Big[(1 - i \alpha) \, \overline \psi \gamma^d \overline D \, \psi - (1 + i  \alpha) \overline{(\overline D \psi)}\, \gamma^d \, \psi \Big],
\end{align}
where $\overline D$ is the gravitational covariant derivative, and $\alpha \in \mathbb R$ is a constant parameter.  The case of constant $\overline \gamma$ has been discussed in \cite{freidel,rovelli} (in fact,  Ref.~\cite{rovelli} deals with minimally-coupled fermions, {\it i.e}. the limit $\alpha = 0$),  
whilst the work of \cite{holstferm} extended the analysis to coordinate-dependent BI, $\overline \gamma (x)$. The extension of the BI to a coordinate dependent quantity implies:  \\
(i) additional terms of interaction of the fermions (F) with the derivative of the BI field $\partial_\mu \overline \gamma$:
\begin{align}\label{fdg}
\mathcal S_{F\, \partial \overline \gamma} = \frac{1}{2} \int \sqrt{-g} \,  \Big( \frac{3}{2 ( \overline \gamma^2 + 1)} \, \partial^\mu \overline \gamma \, \Big[ -J^5_\mu + \alpha \, \overline \gamma (x) \, J_\mu \Big]\Big),
\end{align}
with $J^5_\mu$ the axial current \eqref{axialcurr}, 
and 
\begin{align}\label{vector}
J_\nu = \overline \psi \, \gamma_\nu \, \psi~, 
\end{align}
the vector current.\\
(ii) Interaction terms of fermions with non-derivative $\overline \gamma (x)$ terms:\footnote{A different fermionic action, using non-minimal coupling of fermions with $\gamma^5$, has been proposed in \cite{mercuri}  as a way to resolve an inconsistency of the Holst action, when coupled to fermions, in the case of constant $\gamma$. In that proposal, the $1 + i \alpha$ factor in \eqref{fermionnonder} below, is replaced by the Dirac-self-conjugate quantity 
$1 - i \, \alpha \, \gamma^5$. The decomposition of the torsion into its irreducible components in the presence of the Holst action with arbitrary (constant) BI prameter, leads to an inconsistency, implying that the vector 
component of the torsion is proportional to the axial fermion current, and hence this does not transform properly under improper Lorentz transformations. With the aforementioned modification of the fermion action the problem is solved, as demonstrated in \cite{mercuri}, upon choosing $\alpha = \overline \gamma$, which eliminates the vector component of the torsion. But this inconsistency is valid only if $\overline \gamma$ is considered as a constant. Promotion of the BI parameter $\overline \gamma$ to a {\it pseudoscalar} field, $\overline \gamma (x)$, resolves this issue, as discussed in \cite{holstferm}, given that one obtains in that case consistent results, in the sense that the vector component of the torsion transforms correctly under parity, as a vector, since it contains now, apart from terms proportional to the vector fermionic current \eqref{vector}, also terms proportional to the product of the BI pseudoscalar with the axial fermionic current \eqref{axialcurr}, as well as terms of the form $\overline \gamma \partial_\mu \overline \gamma$, all transforming properly as vectors under improper Lorentz transformations.}
\begin{align}\label{fermionnonder}
\mathcal S_{F -{\rm non-deriv~\overline \gamma}} &= \frac{i}{2} \int \sqrt{-g} \Big[ 
 \Big[(1 - i \alpha) \, \overline \psi \gamma^d D^{\Gamma} \, \psi - (1 + i  \alpha) \overline{(D^\Gamma \psi)}\, \gamma^d \, \psi \Big]  \nonumber \\
 &- \int \sqrt{-g} \, \frac{3}{16(\overline \gamma^2 + 1)} \Big[ J_\mu^5 \, J^{5\mu} - 2\alpha \, \overline \gamma \, J_\mu^5 \, J^{\mu} - \alpha^2 \, J_\mu \, J^\mu \Big],
\end{align}
with $D^\Gamma$ the Riemannian gravitational covariant derivative, expressed in terms of the 
torsion-free Christoffel connection, which is the result of \cite{freidel}, as expected, because this term contains non derivative terms of the BI. 

 We mention at this stage that the generalised four-fermion interactions \eqref{fermionnonder}, which involve {\it attractive} channels among the fermions, may justify (some of) the expectations of \cite{neubert} on the r\^ole of torsion-induced fermion condensates in the early universe cosmology, which cannot characterise the \eqref{4fermi} repulsive terms. Similar features, and thus differences from our string-inspired model, characterise the classical cosmological torsion model of \cite{tors2}, as discussed in the main text, section \ref{sec:string}, where again various self-interactions among properly modified chiral fermions in the model, with various coefficients, are induced as a result of torsion.

We also observe from \eqref{fermionnonder} that the case $\alpha =0$ (minimal coupling), corresponds to a four-fermion axial-current \eqref{4fermi}, which however depends on the BI field. Thus, this limiting theory is not equivalent to our string-inspired model, in which the corresponding quantum-torsion-induced four-fermion axial-current-current interaction \eqref{4fermi} is independent of the KR axion field $b(x)$, although both cases agree with the sign of that interaction.

\end{document}